\documentclass[twocolumn]{aastex63}
\usepackage{lineno}

\received{*, *}
\revised{*, *}
\accepted{*}
\newcommand{\egcite}{\citep[e.g.,][]}

\shorttitle{A Global Collapsing Hub-Filament Cloud G326.611+0.811}
\shortauthors{He et al.}
\graphicspath{{./}{figures/}}

\begin{document}

\title{Investigating a Global Collapsing Hub-Filament Cloud G326.611+0.811}

\correspondingauthor{Yu-Xin He, Hong-Li Liu}
\email{heyuxin@xao.ac.cn, hlliu0104@ynu.edu.cn}

\author[0000-0002-8760-8988]{Yu-Xin He}
\affiliation{Xinjiang Astronomical Observatory, Chinese Academy of Sciences, Urumqi 830011, PR China}
\affiliation{Key Laboratory of Radio Astronomy, Chinese Academy of Sciences, Urumqi 830011, PR China}
\affiliation{Xinjiang Key Laboratory of Radio Astrophysics, Urumqi 830011, PR China}

\author[0000-0003-3343-9645]{Hong-Li Liu}
\affiliation{School of Physics and Astronomy, Yunnan University, Kunming 650091, PR China}

\author[0000-0002-4154-4309]{Xin-Di Tang}
\affiliation{Xinjiang Astronomical Observatory, Chinese Academy of Sciences, Urumqi 830011, PR China}
\affiliation{Key Laboratory of Radio Astronomy, Chinese Academy of Sciences, Urumqi 830011, PR China}

\author[0000-0003-2302-0613]{Sheng-Li Qin}
\affiliation{School of Physics and Astronomy, Yunnan University, Kunming 650091, PR China}

\author[0000-0003-0356-818X]{Jian-Jun Zhou}
\affiliation{Xinjiang Astronomical Observatory, Chinese Academy of Sciences, Urumqi 830011, PR China}
\affiliation{Key Laboratory of Radio Astronomy, Chinese Academy of Sciences, Urumqi 830011, PR China}

\author{Jarken Esimbek}
\affiliation{Xinjiang Astronomical Observatory, Chinese Academy of Sciences, Urumqi 830011, PR China}
\affiliation{Key Laboratory of Radio Astronomy, Chinese Academy of Sciences, Urumqi 830011, PR China}

\author{Si-Rong Pan}
\affiliation{School of Physics and Astronomy, Yunnan University, Kunming 650091, PR China}

\author{Da-Lei Li}
\affiliation{Xinjiang Astronomical Observatory, Chinese Academy of Sciences, Urumqi 830011, PR China}
\affiliation{Key Laboratory of Radio Astronomy, Chinese Academy of Sciences, Urumqi 830011, PR China}

\author[0000-0003-0596-6608]{Meng-Ke Zhao}
\affiliation{Xinjiang Astronomical Observatory, Chinese Academy of Sciences, Urumqi 830011, PR China}
\affiliation{University of Chinese Academy of Sciences, Beijing, 100049, PR China}

\author{Wei-Guang Ji}
\affiliation{Xinjiang Astronomical Observatory, Chinese Academy of Sciences, Urumqi 830011, PR China}
\affiliation{Key Laboratory of Radio Astronomy, Chinese Academy of Sciences, Urumqi 830011, PR China}

\author{Toktarkhan Komesh}
\affiliation{Energetic Cosmos Laboratory, Nazarbayev University, Nur-Sultan 010000, Kazakhstan}
\affiliation{Faculty of Physics and Technology, Al-Farabi Kazakh National University, Almaty, 050040, Kazakhstan}

\begin{abstract}
We present the dynamics study toward the G326.611+0.811 (G326) hub-filament-system (HFS) cloud using the new APEX observations of  both $^{13}$CO and C$^{18}$O (J = 2--1). The G326 HFS cloud constitutes a central hub and at least four hub-composing filaments that are divided into a major branch of filaments (F1, and F2) and a side branch (F3--F5). The cloud holds ongoing high-mass star formation as characterised by  three massive dense clumps (i.e., 370--1100\,$M_{\odot}$ and 0.14--0.16\,g~cm$^{-2}$ for C1--C3) with the high clump-averaged mass infalling rates ($>10^{-3}$\,$M_{\odot}$ yr$^{-1}$) within in the major filament branch, and the associated point sources bright at 70\,$\mu$m typical of young protostars. Along the five filaments, the velocity gradients are found in both $^{13}$CO and C$^{18}$O (J = 2--1) emission, suggesting that the filament-aligned gravitational collapse toward the central hub (i.e., C2) is being at work for high-mass star formation therein. Moreover, a periodic velocity oscillation along the major filament branch is revealed in both $^{13}$CO and C$^{18}$O (J = 2--1) emission with a characteristic wavelength of $\sim$3.5\,pc and an amplitude of $\sim$0.31--0.38\,km s$^{-1}$. We suggest that this pattern of velocity oscillation in G326 could arise from the clump-forming gas motions induced by gravitational instability. Taking into account the prevalent velocity gradients, the fragmentation of the major branch of filaments, and the ongoing collapse of the three massive dense clumps, it is indicative that G326 is a HFS undergoing global collapse.

\end{abstract}


\section{Introduction} \label{sec:intro}

Thanks to \emph{Herschel} observations in the far-infrared and submillimetre regime, the filamentary structures in molecular clouds have been demonstrated to be ubiquitous across the Milky Way, and recognised as a key intermediate bridge between the diffuse interstellar medium (ISM) and dense cores that can further collapse for star formation \citep{2010A&A...518L.102A,2010PASP..122..314M,2011A&A...529L...6A,2014prpl.conf...27A,2016A&A...590A...2S,2018MNRAS.478.2119L,2019MNRAS.487.1259L,2020ApJ...901...31L,2022MNRAS.510.5009L,2019MNRAS.489.4771G,2023A&A...669A.120Z}.
Due to their ubiquity in the ISM, some of filaments can  overlap inevitably with each other, resulting in a special web of filaments that comprise three or more filaments converging together toward the web node.
This special web system is defined as the hub-filament system (HFS hereafter, \citealt{2009ApJ...700.1609M}), and regarded as a unique category of filaments for star formation, especially for high-mass star and star cluster formation \egcite{2009ApJ...700.1609M,2020A&A...642A..87K,2012A&A...543L...3H,2013A&A...555A.112P,2016ApJ...824...31L,2023MNRAS.tmp..140L,2018ApJ...852...12Y,2020ApJ...891...84C,2022A&A...658A.114K,2012A&A...540L..11S}.
In the HFS definition, the web node is defined as the hub while the associated individual filaments as the hub-composing filaments. Generally, the centrally located hub has a lower aspect ratio, but a higher column density than the associated filaments. \egcite{2009ApJ...700.1609M,2020A&A...642A..87K,2023MNRAS.tmp..140L}. In addition, pronounced velocity gradients along the hub-composing filaments have been widely detected \egcite{2013A&A...555A.112P,2016ApJ...824...31L,2018ApJ...852...12Y,2019A&A...629A..81T,2020ApJ...891...84C}, suggesting that filament-rooted gas flows increase the available mass reservoir, which can be transferred to the dense cores forming there (or allow to form more cores building up a cluster).
These observational facts have been considered currently in several major star formation models, such as the global hierarchical collapse \egcite{2019MNRAS.490.3061V}, and clump-fed accretion \citep{2018ARA&A..56...41M}, where the HFSs are often reproduced to be a common signature as the preferential system of cluster and/or high-mass star formation.
Therefore, dedicated, and detailed studies on star-formation related processes (e.g., kinematics, and dynamics) in HFS clouds are crucial to understanding star formation, particularly for high-mass star formation.

The target of this paper is G326.611+0.811 (\emph{RA} = 15$^{\rm h}$43$^{\rm m}$34.00$^{\rm s}$, \emph{Dec} = $-$53$^\circ$57$^\prime$2$^{\prime \prime}$.9, \citealt{2009A&A...505..405P}, hereafter, G326). It is an HFS (see Fig.\,\ref{fig:morphology_of_the_source}), located at a distance of $\sim 2.7 \pm 0.4$\,kpc. This distance is derived from the average distance of three associated dense clumps (AGAL326.626+00.834, AGAL326.607+00.799, and AGAL326.647+00.749), whose distances were determined by \citet{2017AJ....154..140W}. The distance to the G326 cloud, along with its uncertainty ($\sim$15\%), is in agreement with the results of \citet{2020MNRAS.496.3482P}. Hereafter, these three clumps are referred to as C1, C2, and C3, respectively, which can be seen in Fig.\,1. Physical parameters of the ATLASGAL clumps are listed in Table~\ref{tab:atlasgal_clumps}. The mass and mass surface density of the three clumps are in [370, 1100]\,$M_{\odot}$, and [0.14, 0.16]\,g~cm$^{-2}$, respectively. These clumps fall within the region of massive star formation in the mass¨Csize relationship shown in Fig. 24 of \citet{2014MNRAS.443.1555U}. This region is constrained by \citet{2008Natur.451.1082K} and \citet{2010ApJ...723L...7K}, and is also discussed in \citet{2017MNRAS.466.3682B}. These findings lead the authors to suggest that the three clumps most likely form high-mass stars. Moreover, G326 is perhaps at an early evolutionary stage of high-mass star formation (see below). This allows to investigate the initial physical conditions (e.g., kinematics, and dynamics) related to star formation. The major focus of this paper is on the dynamics of the G326 cloud, particularly on the mass accretion signatures from the large-scale filamentary structures down to the smaller-scale clumps, by analysing  the molecular line data of J = 2--1 of $^{13}$CO and C$^{18}$O observed with APEX (see Sect.\,\ref{sec:obs}). The paper is structured as follows.
A brief description of  the APEX observations and the archival Herschel data are given in Section~\ref{sec:obs};  the major results are presented in Section~\ref{sec:res};  the dynamics revealed in G326 is discussed in  Section~\ref{subsec:dis};  finally, the summary is placed in Section~\ref{sec:sum}.

	\begin{deluxetable*}{cccccccccccc} \label{tab:atlasgal_clumps}
		\tablecaption{Physical parameters of the ATLASGAL clumps. The columns are as follows: (1) Clump name; (2) Reference name; (3) Galactic longitude; (4) Galactic latitude; (5) Kinematic distance; (6) Lower range to kinematic distance; (7) Upper range to kinematic distance; (8) Effective radius; (9) Dust temperature; (10) Clump mass derived from the integrated 870 $\mu$m emission; (11) Mass surface density; (12) Virial mass.}
		\tablewidth{0pt}
		\tablehead{ \colhead{Clump name} & \colhead{Reference name} &\colhead{$l$} & \colhead{$b$} & \colhead{$Kd$ \tablenotemark{a}} &
			\colhead{$b\_{Kd}$ \tablenotemark{a}} & \colhead{$B\_{Kd}$\tablenotemark{a}} & \colhead{$R_{eff}$} & \colhead{$T_{d}$} & \colhead{$M$} & \colhead{$\Sigma$} & \colhead{$M_{vir}$}  \\
			& & \colhead{deg} & \colhead{deg} & \colhead{kpc} & \colhead{kpc} & \colhead{kpc} & \colhead{pc} & \colhead{$K$} & \colhead{$M_{\odot}$} & \colhead{$g~cm^{-2}$} & \colhead{$M_{\odot}$} \\
   	  \colhead{(1)} & \colhead{(2)} & \colhead{(3)} & \colhead{(4)} & \colhead{(5)} & \colhead{(6)} & \colhead{(7)} & \colhead{(8)} & \colhead{($9$)} & \colhead{($10$)} & \colhead{($11$)} & \colhead{($12$)} }
		\startdata
		AGAL326.626+00.834 & C1 & 326.626 & +00.834 & 2.72 & 2.28 & 3.13 & 0.42 & 16.1 & 376 & 0.14 & 281\\
		AGAL326.607+00.799 & C2 & 326.607 & +00.799 & 2.66 & 2.22 & 3.08 & 0.69 & 16.8 & 1099 & 0.16 & 370\\
		AGAL326.647+00.749 & C3 & 326.647 & +00.749 & 2.68 & 2.24 & 3.10 & 0.60 & 17.2 & 869 & 0.16 & 467\\
        \enddata
		\tablenotetext{a}{These values are from \citet{2017AJ....154..140W}.}
	\end{deluxetable*}

\section{Observations and data reduction} \label{sec:obs}
\subsection{Molecular line observations}
Molecular line observations of $^{13}$CO (J = 2--1, $\nu$ = 220.39684 GHz) and C$^{18}$O (J = 2--1, $\nu$ = 219.560358 GHz) were made simultaneously with an effective spectral resolution of 114 KHz (e.g. 0.15 km s$^{-1}$) at a tuned central frequency of 220 GHz in the on-the-fly mode on 2017 September 24 using the Atacama Pathfinder Experiment (APEX) 12-m telescope \citep{2006A&A...454L..13G}.
The half power beam width (HPBW) is $\sim$28 arcsec.
The mapping observations were centered at ($\alpha_{\rm J2000.0}$, $\delta_{\rm J2000.0}$) = (15$^{\rm h}$43$^{\rm m}$47\fs23, $-$53\arcdeg55\arcmin59\farcs4) with a rectangular size of 9\,arcmin $\times$ 11\,arcmin.
Calibration was performed using the chopper-wheel technique and the output intensity scale given by the system was $T_{\rm A}^*$, which represents the antenna temperature corrected for atmospheric attenuation.
For further analysis, $T_{\rm A}^*$ was converted to the main beam brightness temperature ($T_{\rm mb}$) {\bf via} $T_{\rm mb}$ = $T_{\rm A}^*$/$\eta_{\rm mb}$, where the main beam efficiency $\eta_{\rm mb}$ is 0.75.
The spectral data were reduced using  the \emph{GILDAS} (Grenoble Image and Line Data Analysis Software) software\footnote{GILDAS is a radio astronomy software developed by IRAM. See {\tt http://www.iram.fr/IRAMFR/GILDAS}}. The reduced data have a typical root mean square (rms) value of $T_{rms}$ $\thickapprox$ 0.08 K.

\subsection{Herschel data}
G326 was observed as part of the \emph{Herschel} infrared Galactic Plane Survey (Hi-GAL) \citep{2010PASP..122..314M}. The Hi-GAL is an unbiased photometric survey of the inner Galactic plane in the five wavebands of 70, 160 $\mu$m \citep[PACS,][]{2010A&A...518L...2P} as well as 250, 350, and 500 $\mu$m \citep[SPIRE,][]{2010A&A...518L...3G} onboard the \emph{Herschel} satellite. The corresponding angular resolutions are 9.2$^{\prime\prime}$, 12.0$^{\prime\prime}$, 17.0$^{\prime\prime}$, 24.0$^{\prime\prime}$, and 35.0$^{\prime\prime}$, respectively, which corresponds to 0.12, 0.16, 0.22, 0.31, and 0.46\,pc at the distance of G326. The detailed descriptions of the pre-processing of the data up to usable high-quality images can be found in \citet{2011MNRAS.416.2932T}.

\section{Results} \label{sec:res}
\begin{figure*}[ht!]
\begin{center}
\includegraphics[scale=1.,angle=0]{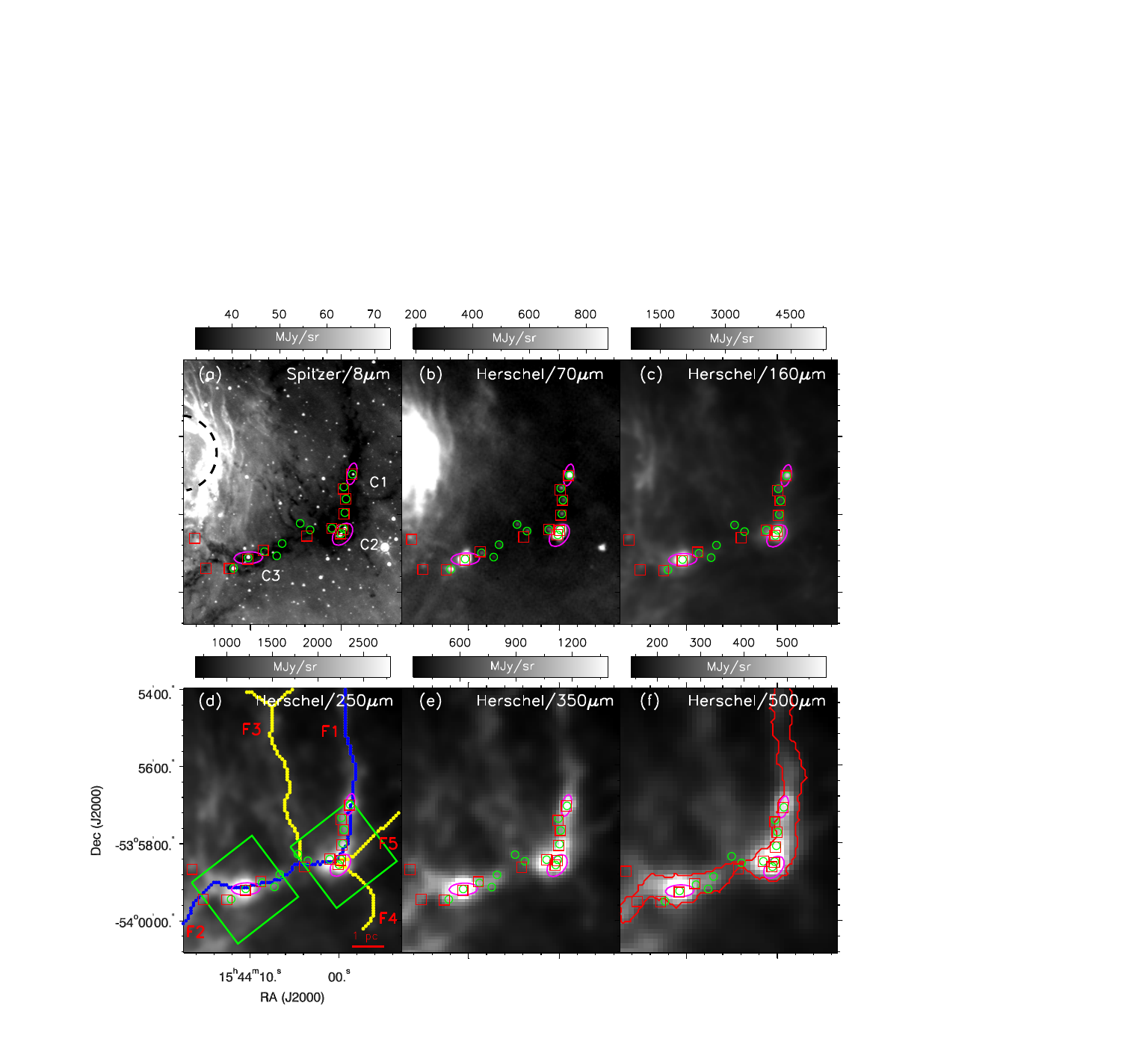}
\caption{Morphology of G326 seen at \emph{Spitzer} 8 $\mu$m, and \emph{Herschel} 70, 160, 250, 350, and 500 $\mu$m. The green circles, magenta ellipses, and red rectangles mark the Herschel 70 $\mu$m point sources, the dust clumps \citep{2014A&A...568A..41U}, and the compact ArT\'eMiS sources \citep{2020MNRAS.496.3482P}, respectively. In panel (a), part of the black dashed circle indicates the infrared bubble MWP1G326725+007745 \citep{2012MNRAS.424.2442S}. The dust clumps are referred to as C1, C2, and C3. In panel\,(d), two green rectangles show the areas covered by the Millimetre Astronomy Legacy Team 90 GHz (MALT90) Survey \citep{2013PASA...30...57J}. The five sub-filaments are referred to as F1, F2, F3, F4, and F5. The major branch of filaments, composed of F1 and F2 (longest and represented as F1+F2 in Table~\ref{tab:sub_filaments}), and branched skeletons are shown as blue and yellow curves, respectively. In panel (f), the outlined sub-region is the masking area in the algorithm \emph{FilFinder} \citep{2015MNRAS.452.3435K} and is used to trace the major branch of filaments.}
\label{fig:morphology_of_the_source}
\end{center}
\end{figure*}

\subsection{Dust properties} \label{sec:dust_properties}
H$_{2}$ column density and dust temperature maps were obtained by simultaneously fitting the intensities at 160, 250, 350, and 500 $\mu$m pixel by pixel using a single temperature gray-body dust emission model \citep{2016ApJ...818...95L,2017A&A...602A..95L,2018MNRAS.478.2119L}. Before the fitting, we made use of the routine \footnote{https://github.com/esoPanda/FTbg/.} developed by \citet{2015MNRAS.450.4043W} to remove foreground/background emission. The 160, 250, and 350 $\mu$m images are convolved and regridded  to the resolution and pixel size of the 500\,$\mu$m image's (i.e., 35.0$^{\prime\prime}$, and 12$^{\prime\prime}$, respectively). For each pixel, the intensities at the four wavelengths were modelled as

\begin{equation}
I_\nu(T_{\rm d},N_{\rm H_2}) \, = \, B_\nu(T_{\rm d})\left(1-e^{-\mu_{\rm H_{2}} m_{\rm H} \kappa_\nu N_{\rm H_2} /{\rm R}}\right),
\label{eq-Idust}
\end{equation}
where $I_\nu$ is the observed intensity, $B_\nu(T_{\rm d})$ is the Planck function at the dust temperature $T_{\rm d}$, $N_{\rm H_2}$ is the column density of molecular hydrogen, $m_{\rm H}$ is the hydrogen atom mass, R is the gas to dust mass ratio (assumed to be 100), $\mu_{\rm H_{2}}$ = 2.8 is the mean molecular weight of the interstellar medium \citep{2008A&A...487..993K}, and $\kappa_\nu$ is the dust absorption coefficient; it follows $\kappa_{\nu} = 5.0(\nu/600~\rm GHz)^{\beta}$, where  $\beta$ = 1.75 is assumed \citep[e.g.,][]{2018ApJ...852...12Y,2019A&A...622A.155Y,2014A&A...564A..45P}. The free parameters are $T_{\rm d}$ and $N_{\rm H_2}$ in the model fitting. According to the error estimation method by \citet{2020MNRAS.492.5420S}, taking into account the systematic errors in mosaic calibration, there is a 20\% uncertainty in the Herschel fluxes at each wavelength band. These uncertainties propagate to the fitted parameters $T_{\rm d}$ and $N_{\rm H_2}$, resulting in an approximate $\sim$9\% uncertainty in these parameters.

\subsection{HFS morphology seen in continuum emission} \label{sec:infrared_dark_hub-filament_cloud}
Fig.\,\ref{fig:morphology_of_the_source} displays the morphology of G326. It appears as an IR dark HFS cloud in Sptizer 8\,$\mu$m emission. The HFS is composed of a central hub bright at 8\,$\mu$m spatially coincided with the C2 clump and the 8\,$\mu$m dark lanes intersecting toward the hub at least from four directions (see Fig.\,\ref{fig:morphology_of_the_source}a). These dark lanes correspond to the five filaments seen in the dust continuum images longward of 250\,$\mu$m well (see Fig.\,\ref{fig:morphology_of_the_source}d--f).
In order to trace these filamentary structures, we use the algorithm \emph{FilFinder}\footnote{{\tt https://github.com/e-koch/FilFinder}} \citep{2015MNRAS.452.3435K} to extract their skeletons from the H$_{2}$ column density ($N_{\rm H_2}$) map. The production of this map can be seen in Sect.~\ref{sec:dust_properties}. Five filaments are identified, referred to as
F1--F5. Their identity  is indicated in the panel\,(d) with the longest skeleton representative of the major branch of filaments (i.e., composed of F1 and F2 in the blue curve and listed as F1+F2 in Table~\ref{tab:sub_filaments})  and the side branch of filaments (F3--F5 in the yellow curves). Four filaments appear to intersect toward the central C2 clump, constituting a hub-filament system.

It is worth noting that  extended emission bright at 8.0 and 70 $\mu$m in the east to the HFS cloud corresponds to an infrared bubble MWP1G326725+007745 \citep{2012MNRAS.424.2442S}. This bubble has been identified as a classical H{\sc\,ii} region at a distance of 1.8\,kpc \citep{2013ApJS..208...11L,2019ApJS..240...24W}. Due to its distance different from that of G326, we assume that this bubble/H{\sc\,ii} region does not have impact on the G326 cloud. We get further the potential effect of the bubble on G326 ruled out in Section~\ref{subsec:origin_of_velocity_gradient}.

Fig.\,\ref{fig:morphology_of_the_source}b presents the Herschel 70 $\mu$m dust continuum image, where 14 point sources from \citet{2016A&A...591A.149M} are indicated in green circles. They all are embedded in the major branch of filaments (i.e., F1 and F2).
The cloud-associated 70\,$\mu$m point sources are generally thought to be young protostars embedded in  clouds \egcite{2016ApJ...818...95L,2017A&A...602A..95L}.
In addition, 9 of 14 sources do not have 8\,$\mu$m counterparts, suggesting the young nature for most of them. This result indicates that G326 is a star-forming cloud still at very early evolutionary stages.

\subsection{CO gas morphology} \label{sec:co_gas_morphology}
\begin{figure}[ht!]
\begin{center}
\includegraphics[scale=0.45,angle=0]{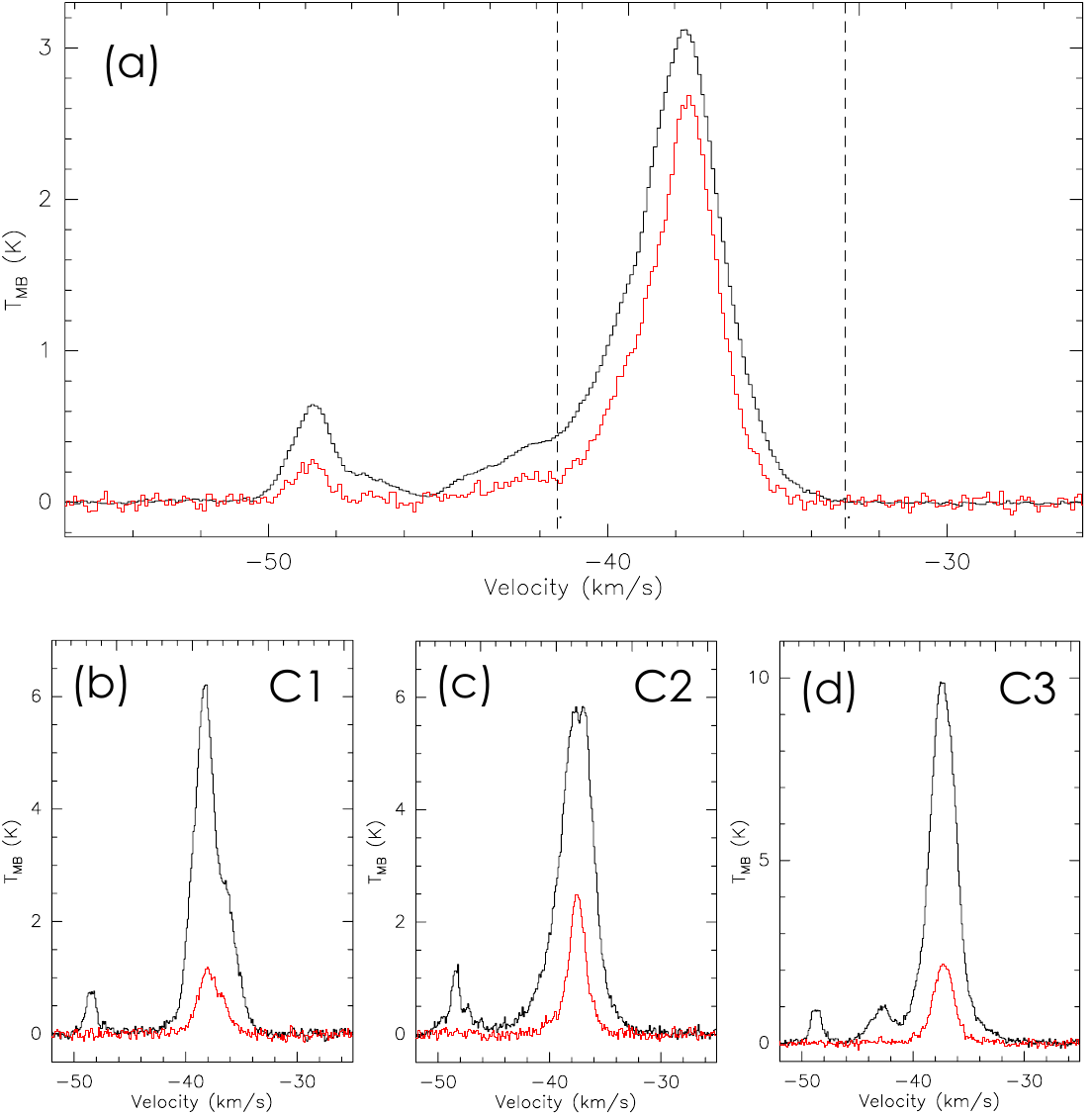}
\caption{(a): Averaged spectra of $^{13}$CO (J = 2--1) (black) and C$^{18}$O (J = 2--1) (red) over the hub-filament cloud of G326. The intensity of C$^{18}$O (J = 2--1) has been reduced by a scale factor of 5 for ease of viewing. The vertical dashed lines show the velocity interval $-$41.5 to $-$33.0 km s$^{-1}$ in which the infrared dark hub-filament cloud of G326 extends. (b): Beam-averaged spectra of $^{13}$CO (J = 2--1) (black) and C$^{18}$O (J = 2--1) (red) for C1.
(c): Same as (b), but for C2. (d): Same as (b), but for C3.}
\label{fig:averaged_spectra}
\end{center}
\end{figure}

\begin{figure*}[ht!]
\begin{center}
\includegraphics[scale=1.2,angle=0]{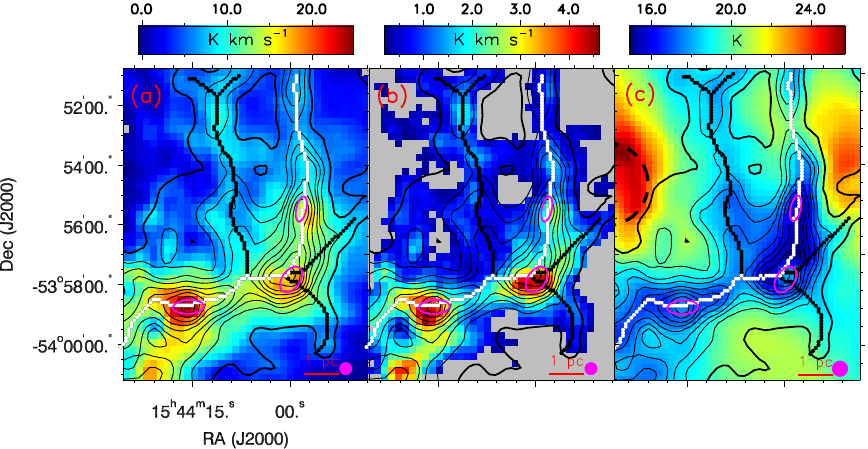}
\caption{(a): Velocity integrated intensity map of $^{13}$CO (J = 2--1) with H$_{2}$ column density from SED fitting overlaid as contours in levels of (0.9, 1.1, 1.3, 1.5, 1.9, 2.4, 3.0, 4.0, 5.0) $\times$ 10$^{22}$ cm$^{-2}$. The velocity interval for integration is [$-$41.5, $-$33.0] km s$^{-1}$. The sigma levels of $^{13}$CO (J = 2--1) integrated map is $\sigma$ = 0.08 K km s$^{-1}$. \textbf{The main and branched skeletons are shown as white and black curves, respectively.} Magenta ellipses mark the dust clumps. The magenta filled circle in the lower right corner indicates the beam size (HPBW) of CO observation. (b): same as panel\,(a), But for C$^{18}$O (J = 2--1). (c): Same as panel\,(a), but for dust temperature. The magenta filled circle in the lower right corner indicates the beam size of \emph{Herschel} 500 $\mu$m waveband.}
\label{fig:13CO_C18O_Td_and_nh}
\end{center}
\end{figure*}

Fig.~\ref{fig:averaged_spectra}\,a shows the averaged spectra of J = 2--1 of $^{13}$CO and C$^{18}$O over the cloud region investigated here. And Fig.~\ref{fig:averaged_spectra}\,b--d show the beam-averaged spectra for C1, C2, and C3. Three velocity components in [$-$51, $-$45]\,km s$^{-1}$, [$-$45, $-$41.5]\,km s$^{-1}$, and [$-$41.5, $-$33.0]\,km s$^{-1}$ are identifiable from the spectra. The careful comparison of the morphologies between the integrated intensities of both species over the three velocity intervals and the continuum emission at multiple wavelengths (Fig.\,\ref{fig:morphology_of_the_source}) suggests that the two weakest velocity components, which are representative of extended diffuse emission, are incongruous with the spatial distribution of filamentary structures in the G326 HFS cloud. They are therefore considered  foreground or background emission unrelated to G326. Fig.~\ref{fig:13CO_C18O_Td_and_nh} shows the spatial distribution of $^{13}$CO  and C$^{18}$O integrated intensities over [$-$41.5, $-$33.0]\,km s$^{-1}$ (i.e., the brightest velocity component) agrees with that of the H$_{2}$ column density map (see below). The systemic velocity, $V_{\rm lsr}$ = $-$37.7\,km s$^{-1}$, is derived from the sine function fit applied to the major filament branch, as presented in Fig.~\ref{fig:velocity_oscillations}. This velocity is consistent with the radial velocities of the clumps used by \citet{2017AJ....154..140W} to calculate the kinematic distance.

Fig.~\ref{fig:13CO_C18O_Td_and_nh} displays the H$_{2}$ column density distribution overlaid on the $^{13}$CO and C$^{18}$O integrated intensity maps (Fig.~\ref{fig:13CO_C18O_Td_and_nh}\,a--b), and the dust temperature map (Fig.~\ref{fig:13CO_C18O_Td_and_nh}\,c). This overall $N_{\rm H_2}$ distribution is anti-correlated with the temperature distributions, suggesting the good performance of our gray-body fitting. In addition, $^{13}$CO shows overall more extended emission than C$^{18}$O and both species have similar emission distribution to the $N_{\rm H_2}$ map, where the HFS morphology of the cloud appears. Three compact clumps are visible in the integrated intensity maps of both $^{13}$CO  and C$^{18}$O. One of them (i.e., C1) seem to slightly offset from the peaks of dust emission, which could be caused by some degree of CO depletion or/and optical depth effects. The average dust temperatures of C1, C2, and C3 are $\sim$16.1, 16.8, and 17.2 K, respectively. Moreover, in the eastern G326 presents relatively higher dust temperatures ($\sim$21.4 K in median), which correspond to the infrared bubble MWP1G326725+007745 (see dashed circle in Fig.\,1a) that does not almost have sufficient gas emissions traced by  $^{13}$CO and C$^{18}$O~(J = 2--1) at the current sensitivity.

\subsection{Velocity gradient} \label{sec:velocity_gradient}
\begin{figure*}[ht!]
\begin{center}
\includegraphics[scale=0.9,angle=0]{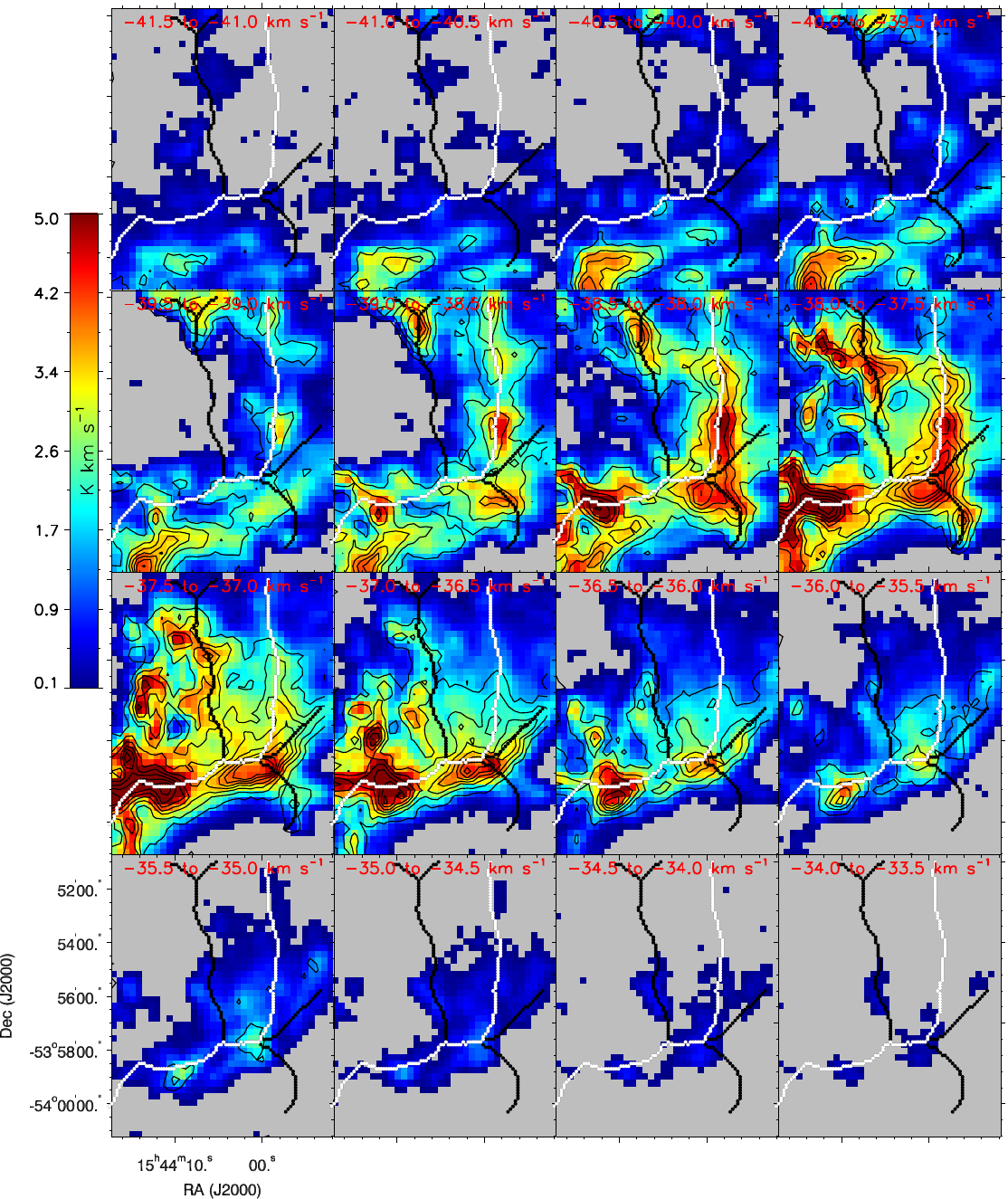}
\caption{Channel maps of $^{13}$CO (J = 2--1) and C$^{18}$O (J = 2--1) lines. $^{13}$CO (J = 2--1) integrated intensity map (start from 6$\sigma$) with C$^{18}$O (J = 2--1) integrated intensity overlaid as contours. The contours start from 6$\sigma$ to the peak value with steps of 8$\sigma$. The $\sigma$ of integrated intensity for both CO isotopologues in the interval velocity range of 0.5 km s$^{-1}$ is 0.018 K km s$^{-1}$. This value is calculated as $T_{rms}\times\sqrt{N_{channels}}\times\delta V$, where $T_{rms}$ = 0.08 K represents the typical rms value of molecular line observations, $N_{channels}$ = 5 denotes the total number of line channels, and $\delta V$ = 0.1 km s$^{-1}$ stands for the velocity resolution. \textbf{The main and branched skeletons are shown as white and black curves, respectively.}}
\label{fig:channel_map_13CO_and_C18O}
\end{center}
\end{figure*}

Fig.~\ref{fig:channel_map_13CO_and_C18O} presents the channel map of $^{13}$CO and C$^{18}$O (J = 2--1) in an increment of 0.5\,km s$^{-1}$ ranging from $-$41.5 to $-$33.5 km s$^{-1}$. From the map, the majority of gas emission is concentrated into velocity [$-$40.5, $-$35.5]\,km\,s$^{-1}$, which agrees with the average spectrum of both species over the G326 cloud. All five filamentary structures, F1 to F5, are detected in both $^{13}$CO and C$^{18}$O emissions. Interestingly, a new filamentary molecular emission branch is observed north of the C3 clump in the velocity range [$-$38.5, $-$35.5]\,km\,s$^{-1}$. This branch is particularly evident in C$^{18}$O. As the velocity increases, the CO emission gradually approaches the C3 clump, showing clear velocity gradients, and likely feeding the C3 clump directly. In the velocity range [$-$41.5, $-$40.5]\,km\,s$^{-1}$, the northern diffused $^{13}$CO emission mainly originates from another velocity component below $-$41.5\,km\,s$^{-1}$ (see Fig.~\ref{fig:averaged_spectra}). Overall, from the channel maps, the emission initially concentrates to the south, and as the velocity increases, the northern emission gradually appears along the five filamentary structures, approaching and feeding the C2 and C3 clumps.

In the velocity range [$-$37.5, $-$36.0]\,km\,s$^{-1}$, most of the detected CO emission is concentrated more to the east. There is also a slight offset of the CO emission distribution from the F3 branch towards the east. The offset CO emission forms a filamentary structure, roughly parallel to the F3 branch, approaching the junction of F3, F1, and F2. This suggests that the material transport along the F3 branch does not directly feed the C2 or C3 clumps but distributes material simultaneously to both C2 and C3 clumps through the junction. Moreover, the northeastward protrusion at the junction of F3, F1, and F2, as indicated by the green vertical solid line in Figs.~\ref{fig:absolute_gradient_of_the_velocity_F1_F5} and \ref{fig:absolute_gradient_of_the_acceleration_F1_F5}, coincides with the highest absolute velocity gradient and the peak of the absolute gradient of acceleration, providing further confirmation of the aforementioned inference. In the low-velocity range of [$-$41.5, $-$39.0] km s$^{-1}$, the molecular gas mainly locates in the northeast and southwest of G326. As the velocity increase from $-$39.0 to $-$37.0 km s$^{-1}$, emissions spatially associated with the main skeleton, while the diffuse gas emissions around the branched skeleton start to appear gradually. In the high-velocity range of [$-$37.0, $-$33.5] km s$^{-1}$, the molecular gas mainly associated with the southwestern two clumps of C2 and C3. Overall, the gas emissions show a gradual shift from the northeastern and southwestern to the lower right dense portion of G326.

\begin{figure*}[ht!]
\begin{center}
\includegraphics[scale=1.2,angle=0]{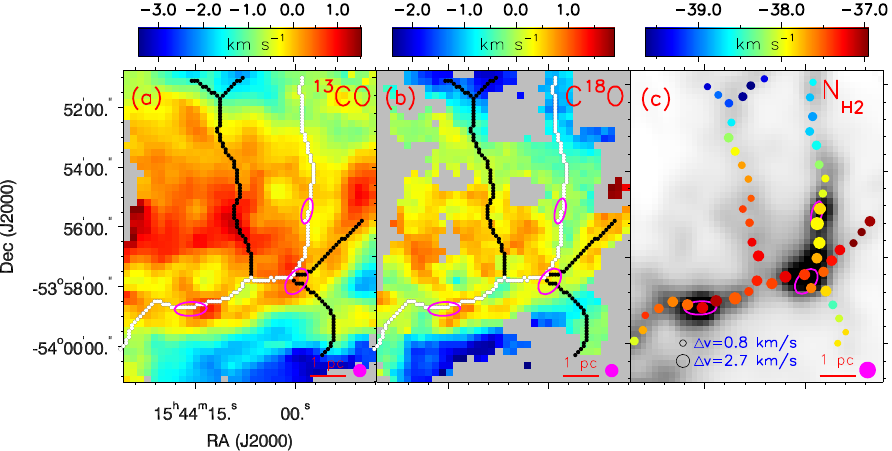}
\caption{(a): The map illustrates the differences between the first velocity moment (velocity field) of $^{13}$CO (J = 2--1) and the systemic velocity of the cloud ($V_{lsr}$ = $-$37.7 km s$^{-1}$). The main and branched skeletons are shown as white and black curves, respectively. Magenta ellipses mark the dust clumps. The magenta filled circle in the lower right corner indicates the beam size of CO observation. (b): Same as left panel, but for C$^{18}$O (J = 2--1). (c): Colour coded line-of-sight velocity centroids of C$^{18}$O (J = 2--1) extracted along filaments overlaid on top of a H$_{2}$ column density map. The size of the symbols indicate the line width of C$^{18}$O (J = 2--1). The magenta filled circle in the lower right corner indicates the beam size of \emph{Herschel} 500 $\mu$m waveband.}
\label{fig:G326_13CO_C18O_and_nh_mom1}
\end{center}
\end{figure*}

\begin{figure}[ht!]
\begin{center}
\includegraphics[scale=0.5,angle=0]{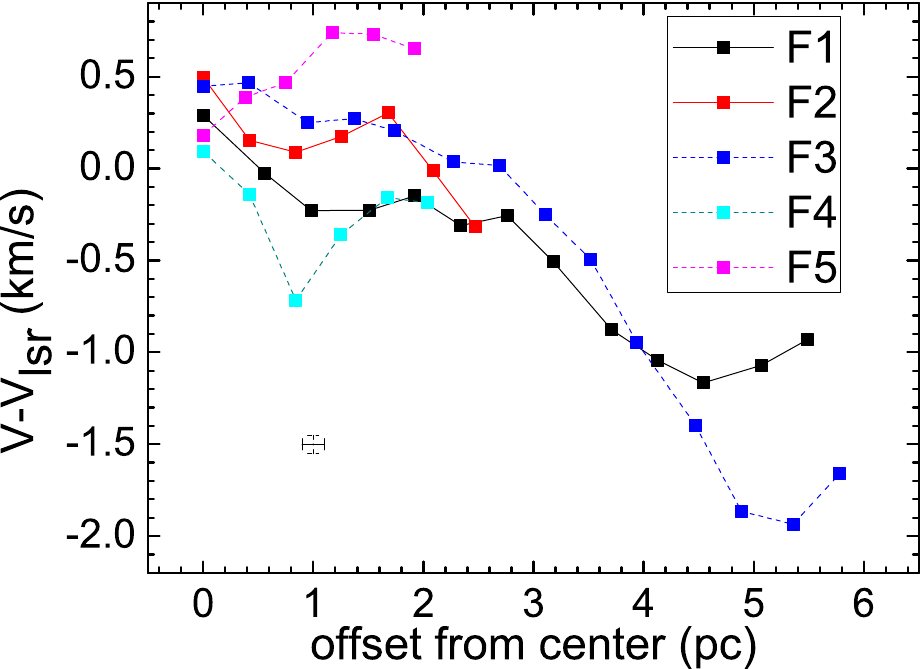}
\caption{The line-of-sight velocity profile of the gas within each filament is plotted as a function of position from the centers of the filaments. The positions are as follows: position C2 for F1 (black solid line), F4 (cyan dotted line), and F5 (magenta dotted line); position at the junction between F3 and the major branch of filaments for F3 (blue dotted line); and position C3 for F2 (red solid line).}
\label{fig:scatterplot_along_the_filament}
\end{center}
\end{figure}

\begin{figure}[ht!]
\begin{center}
\includegraphics[scale=0.6,angle=0]{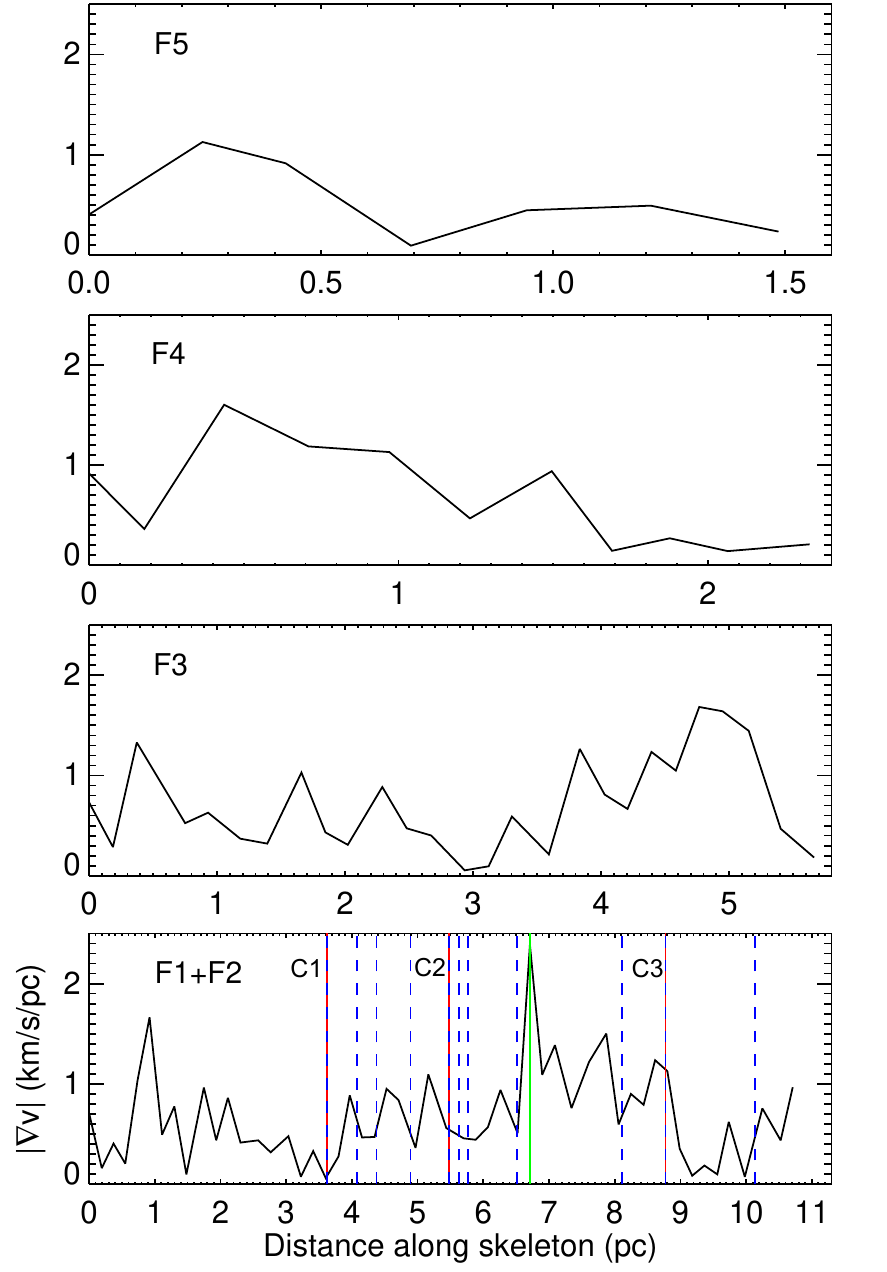}
\caption{The absolute velocity gradient along each skeleton, computed over $\sim$0.19 pc, is plotted as a function of position relative to the centers of the filaments for F3, F4, and F5. The major branch, F1+F2, is plotted from north to southeast. Red vertical solid lines represent the three ATLASGAL clumps. Blue vertical dashed lines indicate the positions of the 11 compact ArT\'eMiS sources situated in the F1+F2 skeleton. The green vertical solid line indicates the position of the junction between F3 and the major branch of filaments.}
\label{fig:absolute_gradient_of_the_velocity_F1_F5}
\end{center}
\end{figure}

\begin{figure}[ht!]
\begin{center}
\includegraphics[scale=0.6,angle=0]{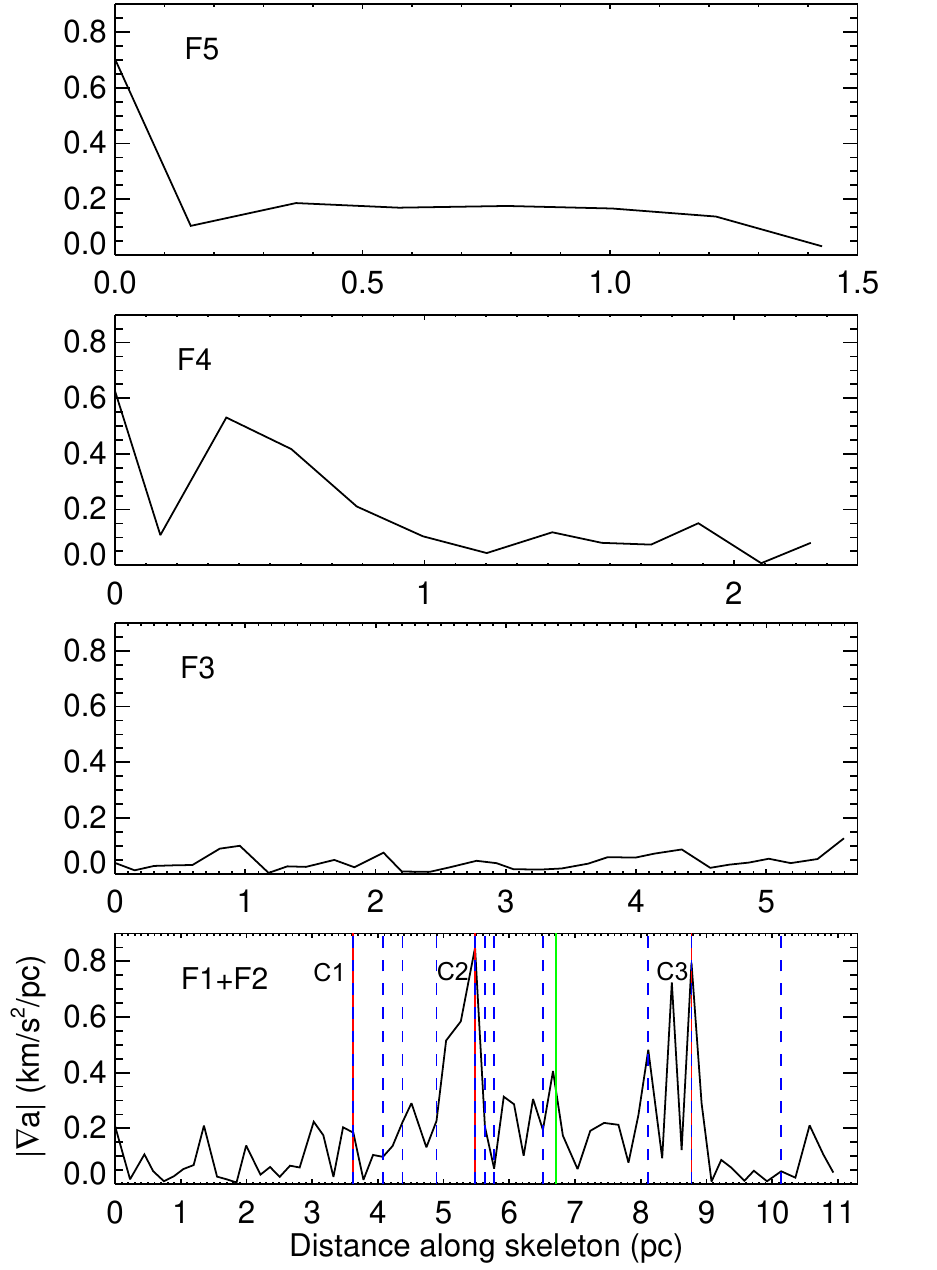}
\caption{Same as Fig.~\ref{fig:absolute_gradient_of_the_velocity_F1_F5}, but for absolute gradient of the acceleration at each position over the size of $\sim$0.15 pc. The units of acceleration gradient are 1$\times$10$^{-12}$ km s$^{-2}$ pc$^{-1}$.}
\label{fig:absolute_gradient_of_the_acceleration_F1_F5}
\end{center}
\end{figure}

The velocity gradients along filaments can be reflected from the map. To better manifest the dynamics of the cloud, we made a map showing the differences between the Moment\,1 maps of the two species and the systemic velocity of the cloud ($V_{lsr}$ = $-$37.7 km s$^{-1}$), as shown in Fig.\,\ref{fig:G326_13CO_C18O_and_nh_mom1}. In Fig.~\ref{fig:G326_13CO_C18O_and_nh_mom1}\,a--b, $^{13}$CO and C$^{18}$O show velocity gradient variations across the entire region. In addition, the filament-rooted line-of-sight velocity gradients mentioned above become more evident in the velocity distribution along the four filaments (see the Fig.~\ref{fig:G326_13CO_C18O_and_nh_mom1}\,c) that are represented by several positions, each sampled every beam (i.e., 28\,\arcsec). The velocity of each position was derived from its beam-averaged spectrum of C$^{18}$O (J = 2--1). Here, the velocity gradients along F1 and F4--F5 appear towards the massive C2 clump, F3 towards the filament junction of F1, F2, and F3, while the one along F2 towards the massive C3 clump. To display the variation of velocity gradients along the filamentary structures more clearly, we have plotted velocity profiles for all five filaments in Fig.~\ref{fig:scatterplot_along_the_filament}. We can see that filamentary structures F1, F3, and F5 initially show a monotonic linear change with increasing distance, and then flatten out as they reach the farthest regions. After excluding the regions with flatten velocity changes, the velocity gradients corresponding to filamentary structures F1, F3, and F5 are measured to be 0.32 $\pm$ 0.05, 0.44 $\pm$ 0.07, and 0.25 $\pm$ 0.04 km s$^{-1}$ pc$^{-1}$, respectively. Overall, filamentary structures F2 and F4 exhibit a decreasing trend with increasing distance, and the corresponding velocity gradients are measured to be 0.33 $\pm$ 0.05 and 0.14 $\pm$ 0.02 km s$^{-1}$ pc$^{-1}$, respectively. Our results are consistent with the findings of \citet{2014A&A...561A..83P} for the SDC13 filamentary molecular cloud at larger scales. The origin of the overall velocity gradient in filamentary structures will be discussed in detail in Sect.\,\ref{subsec:origin_of_velocity_gradient}.

To explore the relationship between the absolute velocity gradient and clump position, we present the absolute value of velocity gradient along the skeletons of F1 to F5 of G326, calculated over an approximate size of $\sim$0.19 pc, plotted against distance along the skeletons in Fig.~\ref{fig:absolute_gradient_of_the_velocity_F1_F5}. Despite the unresolved nature of the compact ArT\'eMiS sources due to relative resolution limitations, our findings indicate that clumps, comprising three ATLASGAL clumps and 11 compact ArT\'eMiS sources, are predominantly situated near peaks or troughs of velocity gradient, where significant velocity changes occur. This pattern resembles the findings of \citet{2018A&A...613A..11W} in the massive filamentary hub SDC13 and \citet{2019A&A...621A.130L} in the high-mass star forming filament G351.776¨C0.527. Different interpretations of this correlation have emerged, including the influence of local motions on nearby clump velocity fields \citep{2019A&A...621A.130L}, indication of accretion from the parent filament \citep{2018A&A...613A..11W}, and the presence of transient motions generated by the turbulent field \citep[detailed in][]{2016MNRAS.455.3640S}. To further analyze the potential causes of the aforementioned velocity gradients, we followed the acceleration amplitude calculation formula proposed by \citet{2018A&A...613A..11W}: $|\vec{a_j}|= \left|-\sum_{i\neq j}\frac{Gm_i}{r_{i,j}^3}\vec{r_{i,j}}\right|$. Here, $i$ and $j$ represent given pixels in the column density map of the G326 source, $m_{i}$ is the mass corresponding to pixel $i$, and $r_{i,j}$ is the distance between pixels $i$ and $j$. Using this formula, we calculated the absolute acceleration gradient variations along the five filamentary structures as a function of distance, as shown in Fig.~\ref{fig:absolute_gradient_of_the_acceleration_F1_F5}.

\begin{figure*}[ht!]
\begin{center}
\includegraphics[scale=1,angle=0]{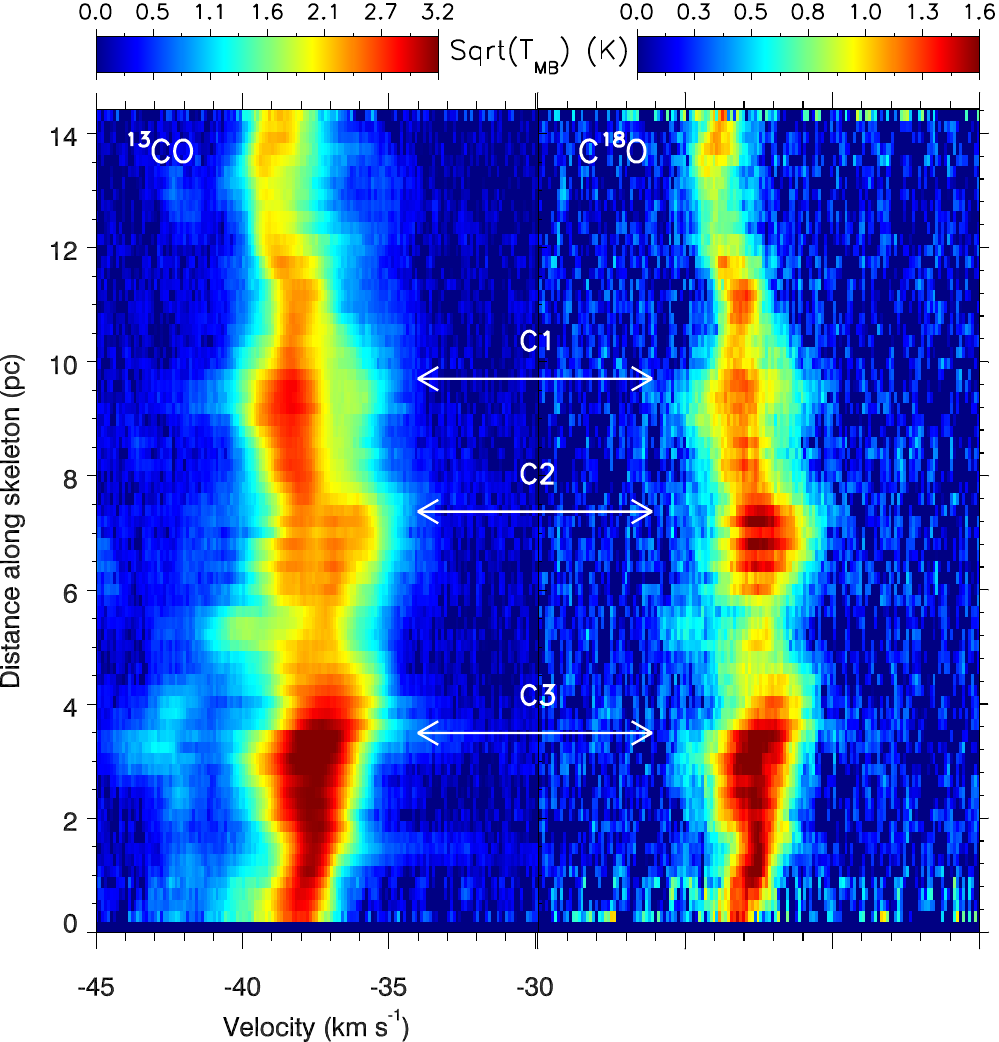}
\caption{Left: $^{13}$CO position-velocity diagram as a function of the distance along the main skeleton (see Fig.~\ref{fig:G326_13CO_C18O_and_nh_mom1}) from the southern to northern end. The color scale is square. The white arrows represent positions of C1, C2, and C3. Right: same as left panel, but for C$^{18}$O.}
\label{fig:13CO_and_C18O_PV_diagram}
\end{center}
\end{figure*}

From the bottom panel of Fig.~\ref{fig:absolute_gradient_of_the_acceleration_F1_F5}, it is evident that the acceleration gradients are associated with the positions of the three ATLASGAL clumps. In the upper two panels of Fig.~\ref{fig:absolute_gradient_of_the_acceleration_F1_F5}, we also observe that the ends of filamentary structures F4 and F5, which are close to clump C2, display the highest absolute acceleration gradients. This indicates that the peak positions of absolute acceleration gradients align with material accumulation points, which are indicative of clump formation locations. Except for the three ArT\'eMiS sources that overlap with ATLASGAL clumps, the remaining eight sources do not exhibit consistent peak correspondences on the absolute acceleration gradient map. This is primarily attributed to the observational resolution limitations of our dataset in this study. Additionally, the gravitational acceleration from clumps leads to noticeable variations in velocity gradients near the clumps, aligning well with the explanations provided by \citet{2019A&A...621A.130L} and \citet{2018A&A...613A..11W}. However, through the above analysis of the filamentary structure G326, we are still unable to determine which interpretation is dominant. Further detailed and larger-sample statistical research is needed to answer this question. Furthermore, the changes in velocity gradients near the clumps manifest as velocity fluctuations in the velocity profiles of the filamentary structures. In addition, on top of these filament-aligned velocity gradients, we find a periodic velocity fluctuation along the major branch of filaments (composed of F1 and F2) as evidenced by the position-velocity diagram of both $^{13}$CO and C$^{18}$O (J = 2--1) along it (see Fig.\,\ref{fig:13CO_and_C18O_PV_diagram}), which will be discussed in more detail in Sect.\,\ref{subsec:velocity_oscillations}.

\subsection{Stability of filaments} \label{sec:stability_of_filaments}
As mentioned earlier, three massive clumps (i.e., C1--C3) and twelve compact ArT\'eMiS sources are embedded in the major branch of HFS that constitutes F1 and F2, suggestive of fragmentation in them. Especially the ArT\'eMiS clumps, which were identified from high spatial resolution ($\sim$8 arcsec) observations by \citet{2020MNRAS.496.3482P}, although our CO molecular line observations have a relatively lower resolution ($\sim$28 arcsec) and cannot resolve these clumps, therefore cannot provide velocity information for these clumps, they still serve as excellent samples for studying filament fragmentation. The average spacing of the 12 ArT'¨¦MiS clumps observed, excluding the one located near the infrared bubble MWP1G326725+007745 in the bottom left corner of Fig.~\ref{sec:res}, is approximately 0.6 pc at a distance of 2.7 kpc. According to the theory initially proposed by \citet{1953ApJ...118..116C} and further improved by \citet{1987PThPh..77..635N} on the self-gravitating fluid cylindrical "sausage" instability model, the cores formed through the fragmentation of highly filamentary clouds will be distributed with an approximate characteristic spacing, which is equal to the wavelength of the fastest growing unstable mode of the fluid instability \citep{2010ApJ...719L.185J}. In the case of an incompressible fluid, the wavelength  is given by $\lambda_{\rm max}$ = 11$R$, where $R$ represents the radius of the cylinder \citep{1953ApJ...118..116C}. In the context of an infinite isothermal gas cylinder, the relationship becomes $\lambda_{\rm max}$ = 22$H$, where $H$ is the isothermal scale height given by $H = c_{\rm s}(4\pi G \rho_{\rm c})^{-1/2}$, with $c_{\rm s}$ representing the sound speed, $G$ being the gravitational constant, and $\rho_{\rm c}$ as the gas mass density at the center of the filament \citep{1987PThPh..77..635N,1992ApJ...388..392I}. Furthermore, the spacing depends on the ratio between the cylinder radius and the isothermal scale height, $R/H$. When $R$ is much greater than $H$ ($R \gg H$), the core spacing approaches that of an infinite radius cylinder, $\lambda_{\rm max}$ = 22$H$. However, for cases where R is much smaller than $H$ ($R \ll H$), the spacing reduces to that of an incompressible fluid, $\lambda_{\rm max}$ = 11$R$.

We will compare the observational results of G326 with the "sausage" instability theory under the assumption that G326 is well approximated by an isothermal cylinder. The radius $R$ = 0.4 pc of G326 is estimated from the major filament branch as outlined by a level of $1.0\times10^{22}$\,cm$^{-2}$ (see Fig.\,\ref{fig:morphology_of_the_source}\,f). In the major branch of filaments, the dust temperature ranges from 15.3 to 18.1 K, with a mean value of 16.9 $\pm$ 0.7 K. This corresponds to a sound speed of $c_{\rm s}$ = 0.25 $\pm$ 0.01 km s$^{-1}$. The averaged gas mass density at the center of the filament is $\rho_{\rm c}$ = 1.06 $\times$ 10$^{-13}$ g cm$^{-2}$, then the isothermal scale height is $H$ = 0.03 pc. Consequently, G326 would fall into the regime where $R \ll H$, and the theoretical spacing between cores should be $\lambda_{\rm max}$ = 22$H$ $\sim$ 0.6 pc. This is in good agreement with the observed spacing of 0.6 pc. This is in good agreement with the observed clump spacing of 0.6 pc, further supporting the model of self-gravitating fluid cylinders fragmenting due to "sausage" instability.

Additionally, it is necessary to assess the gravitational stability of these density structures (i.e., F1, and F2 and their embedded clumps). To this end, we analysed the virial (or critical) line mass of the filaments ($M_{\rm line,vir}$) and virial mass of three clumps ($M_{\rm vir}$). The major branch of filaments investigated here has a total length of $\sim$11.3\,pc. By integrating the H$_{2}$ column density over the major filament branch as outlined by a level of $1.0\times10^{22}$\,cm$^{-2}$ (see Fig.\,\ref{fig:morphology_of_the_source}\,f), we obtained a total mass of  $\sim5\times10^3$\,M$_{\odot}$ with an uncertainty of about 31\%, which arises from the $\sim$9\% uncertainty in $N_{\rm H_2}$ and the $\sim$15\% uncertainty in distance. This mass calculation includes more extended structures than observed in \citet{2020MNRAS.496.3482P}, resulting in a slightly higher mass than the 3260 $M_{\odot}$ determined by \citet{2020MNRAS.496.3482P} and 1151 $M_{\odot}$ determined by \citet{2016A&A...591A...5L}. Taking into account the uncertainties introduced by various parameters, the final mass determination for the major structure of the G326 filament is consistent within the error range. This mass accounts for approximately 40\% of the total mass of the G326, which is $\sim1.2\times10^4$\,M$_{\odot}$. Accordingly, the line mass of the major branch of filaments is $M_{\rm line}=428$\,M$_{\odot}$\,pc$^{-1}$ with an uncertainty of about 35\%, which is a combination of the $\sim$9\% uncertainty from $N_{\rm H_2}$ and the $\sim$31\% uncertainty from the mass. Note that this value is an upper limit since the actual inclination of the filament, which is unknown, was not considered in the estimate.

Following \citet{2019ApJ...877....1D}, $M_{\rm line,vir}$ was calculated as below:
\begin{equation}
M_{\rm line,vir} = \left[1 + \left(\frac{\sigma_{\rm NT}}{c_{\rm s}}\right)^2 \right] \times \left[16~M_{\odot}~pc^{-1} \times \left(\frac{T}{10~K}\right) \right]
\label{equa:virial_line_mass}
\end{equation}
where $T$ is the gas kinetic temperature, $c_{\rm s}$ = $\sqrt{k_{\rm B}T/\mu m_{\rm H}}$ is the sound speed ($\mu$ = 2.33 is the mean molecular weight), and $\sigma_{\rm NT}$ is the non-thermal velocity dispersion, which is defined as follows \citep{2019MNRAS.487.1259L,2017A&A...598A..30T}:
\begin{equation}
\sigma_{\rm NT} = \sqrt{\frac{\Delta v^2}{8\ln 2}-\frac{k_{\rm B} T}{m_{\rm mol}}},
\label{equa:nontherma_lvelocity_dispersion}
\end{equation}
where $m_{\rm mol}$ is the molecular mass (30$m_{\rm H}$ for C$^{18}$O), and $k_{\rm B}$ is the Boltzmann constant. To account for the influence of clump spectral line width (FWHM) on the estimation of filamentary structure linewidth, we excluded the FWHMs within the clump position regions (C1-C3) of the main filamentary structure (outlined in red in Fig.\,\ref{fig:morphology_of_the_source}f). Using the C$^{18}$O FWHM map shown in Fig.\,\ref{fig:FWHM_velocity_width_distribution}, we computed the average FWHM for the remaining extended filamentary region, resulting in an estimated value of $\Delta v$ = 1.79 $\pm$ 0.03 km s$^{-1}$. Assuming local thermodynamic equilibrium (LTE) conditions, the gas kinetic temperature $T$ was treated as the dust temperature $T_{\rm d}$, i.e., $T$ = $T_{\rm d}$. In the major branch of filaments,the non-thermal velocity dispersion of $\sigma_{\rm NT}$ = 1.00 $\pm$ 0.2 km s$^{-1}$. Accordingly, the Mach number $\cal M$ (i.e., ratio of $\sigma_{\rm NT}$/$c_{\rm s}$) was computed to be 4.0 typical of supersonic motions. Finally, the critical line mass was estimated to be $M_{\rm line,vir}$ = 286 $\pm$ 19 M$_{\odot}$ pc$^{-1}$ with the error mostly arising from the uncertainty of the temperature. It turns out that the observed $M_{\rm line}$ is larger than the critical $M_{\rm line,vir}$. This result suggests that G326 would undergo gravitational fragmentation into prestellar cores. Furthermore, the presence of fourteen 70 $\mu$m point sources (Fig.~\ref{fig:morphology_of_the_source}) embedded in the major branch of filaments also indicates that the process of fragmentation is going on in the cloud. One should also note here that the observed $M_{\rm line}$ values were underestimated due to the lack of accounting for the inclination of the G326 cloud. For instance, an inclination angle of 45\degr\ would increase $M_{\rm line}$ by a factor of 1.4. Additionally, the critical $M_{\rm line,vir}$ values may have been overestimated due to no accounting of the external environmental pressure imposed onto the major branch of the filaments

Following the same method as described above, we performed individual calculations for each filament. We found that the masses of the F1 (length of 6.6 pc) and F2 (length of 4.7 pc) components, which constitute the major filament, exceed 2000 $M_{\odot}$, accounting for approximately 23\% and 17\% of the total mass of the G326, respectively. Their line masses are 419 $M_{\odot}~pc^{-1}$ and 440 $M_{\odot}~pc^{-1}$, respectively, while their critical masses are 268 $M_{\odot}~pc^{-1}$ and 306 $M_{\odot}~pc^{-1}$. This indicates that F1 and F2 are undergoing gravitational fragmentation, consistent with the findings of the above major filament study. The intersecting F3, with a length of 5.8 pc, has an estimated mass of approximately 1500 $M_{\odot}$. From Fig.~\ref{fig:G326_13CO_C18O_and_nh_mom1}, it is evident that F3 exhibits a complex velocity gradient, which is reflected in a larger C$^{18}$O linewidth of $\Delta v$ = 2.71 $\pm$ 0.03 km s$^{-1}$. The line mass of F3 is 258 $M_{\odot}~pc^{-1}$, which is lower than its critical line mass of $M_{\rm line,vir}$ = 623 $M_{\odot}~pc^{-1}$, suggesting that it is in a state of gravitational equilibrium. Additionally, we did not detect any ATLASGAL and ArT\'eMiS dense clumps in F3, further indicating that gravitational fragmentation has not yet commenced. Finally, the two shortest filamentary structures, F4 (2.1 pc) and F5 (1.5 pc), intersect with F1 and converge at C2. Their line masses are 136 $M_{\odot}~pc^{-1}$ and 175 $M_{\odot}~pc^{-1}$, respectively, which are lower than their critical line masses of 181 $M_{\odot}~pc^{-1}$ and 381 $M_{\odot}~pc^{-1}$. Similar to F3, they are also in a state of gravitational equilibrium. The total mass of filamentary structures F3-F5 is close to the mass of F2, accounting for approximately 17\% of the entire mass of the G326. These filaments are named F1 to F5 and F1+F2, and their physical parameters are listed in Table~\ref{tab:sub_filaments}.

	\begin{deluxetable}{ccccccc} \label{tab:sub_filaments}
		\tablecaption{Properties of filaments. The columns are as follows: (1) Filament name; (2) Filament length; (3) Dust temperature; (4) Line width of C$^{18}$O; (6) Filament mass; (7) Line mass of the filament; (8) Critical line mass of the filament.}
		\tablewidth{0pt}
		\tablehead{ \colhead{Name} & \colhead{L} &\colhead{$T_{d}$} & \colhead{$\Delta v$} &
			 \colhead{$M$} & \colhead{$M_{line}$} & \colhead{$M_{line,vir}$}  \\
			& \colhead{$pc$}  & \colhead{$K$} & \colhead{} & \colhead{$M_{\odot}$} & \colhead{$M_{\odot}~pc^{-1}$} & \colhead{$M_{\odot}~pc^{-1}$} \\
   	  \colhead{(1)} & \colhead{(2)} & \colhead{(3)} & \colhead{(4)} & \colhead{(5)} & \colhead{(6)} & \colhead{(7)} }
		\startdata
		F1    & 6.6 & 16.7 & 1.73  & 2767 & 419 & 268 \\
		F2    & 4.7 & 17.2 & 1.85  & 2073 & 440 & 306 \\
		F3    & 5.8 & 18.0 & 2.71  & 1493 & 258 & 623 \\
        F4    & 2.1 & 18.5 & 1.37  & 287 & 136 & 181 \\
        F5    & 1.5 & 17.8 & 2.09  & 269 & 175 & 381 \\
        F1+F2 & 11.3 & 16.9 & 1.79  & 4841 & 428 & 286 \\
        \enddata
	\end{deluxetable}

\subsection{Stability of clumps} \label{sec:stability_of_clumps}
Following \citet{2010ApJS..188..313W}, the virial mass of the dense clumps are calculated by
\begin{equation}
M_{\rm vir}(R_{\rm eff}) = \frac{5R_{\rm eff}\Delta v^{2}}{8a_{1}a_{2}Gln2} \approx 209\frac{(\frac{R_{\rm eff}}{1~pc})(\frac{\Delta v}{1~km~ s^{-1}})^{2}}{a_{1}a_{2}} M_{\odot}
\end{equation}
\begin{equation}
a_{1}=\frac{1-\frac{p}{3}}{1-\frac{2p}{5}}, \ p < 2.5,
\end{equation}
where $G$ is the gravitational constant, $R_{\rm eff}$ is the effective radius, $\Delta v$ is the line width, $a_{1}$ is the correction for a power-law density distribution, and $a_{2}$ is the correction for a nonspherical shape \citep{1992ApJ...395..140B}. We assume the three dense clumps are spherical, $a_{2}$ $\approx$ 1. The average $p$ value of 1.77 is used to calculate $a_{1}$ for massive dense clumps \citep{2002ApJS..143..469M}.

The clump masses ($M_{\rm clump}$) of C1, C2, and C3 were estimated to be 376, 1099, and 869 $M_{\odot}$ by integrating the column density $N_{\rm H_2}$ over regions with effective radius of 0.42 $\pm$ 0.07, 0.69 $\pm$ 0.11, and 0.60 $\pm$ 0.10 pc, respectively. The effective radius are calculated from the angular sizes measured by \citet{2014A&A...568A..41U}. \citet{2017MNRAS.471..100E} have carried out mass calculations for the three clumps. However, it's worth noting that only the distance used for calculating the mass of clump C3 is close to 2.7 kpc, with a resulting mass of 390 $M_{\odot}$. This value is lower than our calculated mass. This discrepancy arises because our mass calculations consider the radiation region within the effective radius of the clumps \citep[46 arcsec,][]{2014A&A...568A..41U}, covering a larger area compared to the region chosen by \citet{2017MNRAS.471..100E} for clump C3. We adopted this methodology to facilitate a direct comparison with the virial mass results.

The optically thin N$_{2}$H$^{+}$(J = 1--0) line from the MALT90 survey was used to calculate the virial mass. The derived line width from the hyperfine structure (hfs) fitting using the hfs method in \emph{CLASS} to the clump-average spectrum are 1.77 $\pm$ 0.03\,km s$^{-1}$ for C1, 2.21 $\pm$ 0.01\,km s$^{-1}$ for C2, and 2.42 $\pm$ 0.02 km s$^{-1}$ for C3. In turn, the corresponding virial masses are 194 $\pm$ 50, 500 $\pm$ 94, and 521 $\pm$ 105 $M_{\odot}$, respectively. The uncertainties of the clumps mass and virial mass mostly depend on the uncertainties of distance. The most massive C2 clump is located in the hub center of the HFS cloud (see Fig.~\ref{fig:13CO_C18O_Td_and_nh}), and has the largest virial ratio of $M_{\rm clump}$/$M_{\rm vir}$ = 2.2. The other two dense clumps (i.e., C1, and C3) have virial ratios of 1.9 and 1.7, respectively. We therefore suggest that these clumps are gravitationally unstable, and collapsing through gravitational instability  (see Sect.\,\ref{subsec:global_collapse_of_the_cloud} for details). According to the threshold for massive star formation  ($M(r)\geq870M_{\odot}(r/pc)^{1.33}$, \citealt{2010ApJ...723L...7K}), all of the three clumps have masses above the threshold, and thus are potential sites of high-mass star formation (see Sect.\,\ref{subsec:global_collapse_of_the_cloud}).

\section{Discussion} \label{subsec:dis}

\subsection{Origin of filament-aligned velocity gradient} \label{subsec:origin_of_velocity_gradient}
As mentioned in Section~\ref{sec:velocity_gradient}, filament-aligned velocity gradients are observed along all filaments (i.e., F1--F5).  Due to the proximity in projection between the bubble and the G326 cloud, it is natural to ask if the bubble causes or not the filament-aligned velocity gradients observed here. The infrared bubble is driven, and ionized by an embedded H{\sc\,ii} region \citep[i.e., G326.721+00.773,][]{2019ApJS..240...24W}. In the ionizing process, some molecules like C$^{18}$O are likely to be selectively dissociated by the FUV radiation, even in a deep molecular cloud interior (e.g., \citealt{2014A&A...564A..68S}). For the C$^{18}$O molecule, the selective dissociation can lead to a higher abundance ratio between $^{13}$CO and C$^{18}$O (X$^{13/18}$). The spatial distribution of X$^{13/18}$ would therefore help understand the potential effect of the  bubble on the cloud dynamics.

\begin{figure*}[ht!]
\begin{center}
\includegraphics[scale=1.4,angle=0]{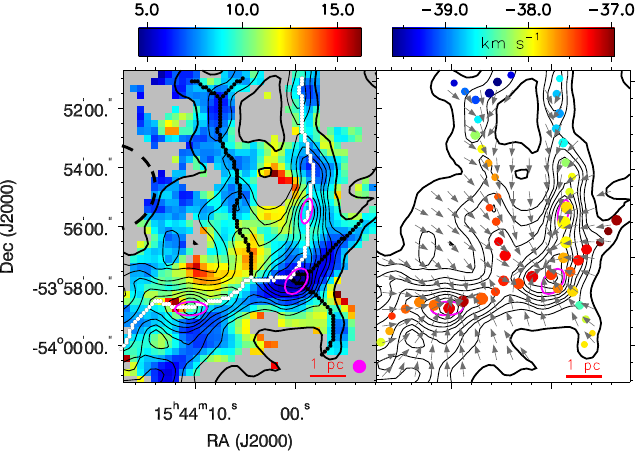}
\caption{Left: Maps of the $^{13}$CO/C$^{18}$O abundance ratio obtained from the $^{13}$CO, and C$^{18}$O J = 2--1 line assuming local thermodynamic equilibrium (LTE). Contours are the H$_{2}$ column density in levels of (0.9, 1.1, 1.3, 1.5, 1.9, 2.4, 3.0, 4.0, 5.0) $\times$ 10$^{22}$ cm$^{-2}$. Magenta ellipses mark the dust clumps. Part of the black dashed circle indicates the infrared bubble MWP1G326725+007745. The main and branched skeletons are shown as white and black curves, respectively. The beam size of CO observation is shown in the lower right corner. Magenta ellipses mark the dust clumps. Right: Projected local gravitational field (gray arrows) overlaid on the H$_{2}$ column density contours in levels of (0.9, 1.1, 1.3, 1.5, 1.9, 2.4, 3.0, 4.0, 5.0) $\times$ 10$^{22}$ cm$^{-2}$ and the colour coded velocity centroids of C$^{18}$O (J = 2--1). The size of the symbols indicate the line width of C$^{18}$O (J = 2--1). The gravitational force vectors are displayed with uniform lengths to emphasize their directions.}
\label{fig:CO_abundance_ratio}
\end{center}
\end{figure*}

Assuming LTE conditions, we estimated the column densities of $N_{\rm ^{13}CO}$ and $N_{\rm C^{18}O}$, and the abundance ratio X$^{13/18}$ = $N_{\rm ^{13}CO}$/$N_{\rm C^{18}O}$. Details of the estimates are given in Appendix~\ref{sec:CO_column_densities}. The obtained abundance ratio X$^{13/18}$ is in [4.66, 19.91] with a mean value of 9.18 $\pm$ 2.07 over the cloud as shown in Fig.~\ref{fig:CO_abundance_ratio}. As a result, the region closed in projection to the bubble does not show significantly different X$^{13/18}$ distribution than those relatively far from the bubble. Quantitatively, the abundance ratio X$^{13/18}$ around the region around the bubble is measured to be  $\sim$7.5, which agrees with the value 7.4 derived from the experiential relation between atomic ratios and the galactocentric distance summarised in \citet{1994ARA&A..32..191W} (a galactocentric distance of 6.1\,kpc adopted here). These results favor that  the bubble is not physically associated the G326 cloud, and thus has  no effect on the cloud dynamics (e.g., the filament-aligned velocity gradients), which agree with the distance estimates of the two objects (i.e., 1.8\,kpc for the bubble versus 2.7\,kpc for the cloud).

\citet{2014A&A...561A..83P} proposed three optional physical processes responsible for the filament-aligned velocity gradients, including the rotation, local gas collision, and gravitational collapse of filament. The rotation can be ruled out and otherwise would tear apart the filaments \citep{2014A&A...561A..83P}. Although the possibility of the local gas collision can not be excluded, the gravitational collapse of filaments toward
the gravitational potential well centers (i.e., the C2 and C3 massive clumps) is the probable interpretation.
Following the deriving method of the local projected gravitational force by \citet{2020ApJ...905..158W}, we estimated the  gravitational vector field from the H${_2}$ column density map, as presented in the right panel of Fig.~\ref{fig:CO_abundance_ratio}. From the figure, the global trend is that the gravitational vectors tend to
point toward the massive C2 and C3 clumps either along the filaments or through the directions slightly inclined to the filaments. This result suggests that gravitational collapse of the filaments could dominate the dynamics of the cloud, which is most likely responsible for the filament-aligned velocity gradients. In addition, the increasing trend of the observed line widths along the filaments toward the massive C2 and C3 clumps can be explained well if the turbulence is converted by the gravitational energy from the outside of the cloud to the gravitational potential well centres \citep{2014A&A...561A..83P}.

\subsection{Velocity oscillations} \label{subsec:velocity_oscillations}
\begin{figure*}[ht!]
\begin{center}
\includegraphics[scale=1.0,angle=0]{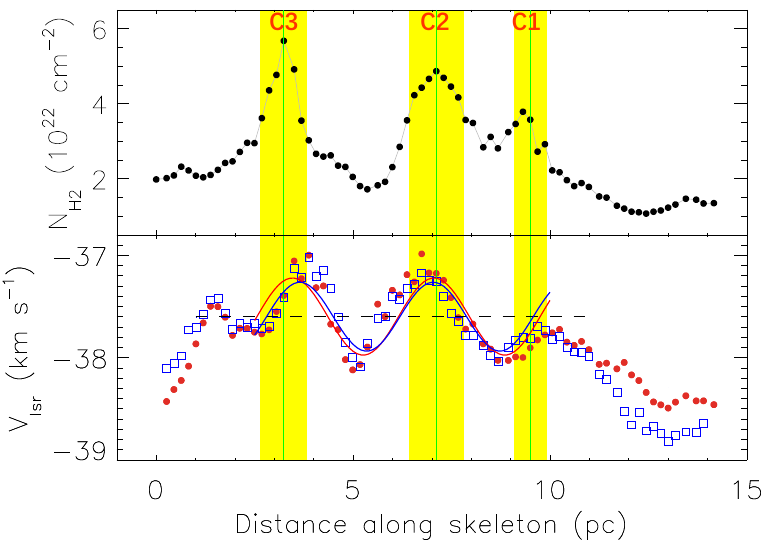}
\caption{Velocity and H$_{2}$ column density distributions extracted along the main skeleton of the filament. Top panel: Density distribution. Density peaks related to the three dense clumps of C1, C2, and C3 are indicated by vertical green solid lines with yellow shadows. The width of each yellow shadow is the size of clump. Bottom panel: The red filled points are $^{13}$CO (J = 2--1) velocities, and fitted with a sine function with a wavelength of $\sim$3.6 pc and an amplitude of $\sim$0.38 km s$^{-1}$. The fitting curve is shown as the red solid line. The blue open squares are C$^{18}$O (J = 2--1) velocities, and fitted with a sine function with a wavelength of $\sim$3.4 pc and an amplitude of $\sim$0.31 km s$^{-1}$. The fitting curve is shown as the blue solid line. The vertical shift of the sine function, representing the systemic velocity, is indicated by a horizontal dashed line at $V_{\rm lsr}$ = $-$37.7 km/s.}
\label{fig:velocity_oscillations}
\end{center}
\end{figure*}

As mentioned earlier, the large-scale velocity oscillation along the major branch of filaments (i.e., F1 and F2)
can be inferred from the position velocity diagram along the branch.  In Fig.~\ref{fig:velocity_oscillations}, we present the distribution of two parameters as a function of position along the major branch of filaments, including the H$_{2}$ column density ($N_{\rm H_2}$, see the top panel), and the velocity centroid of both $^{13}$CO (red points in the bottom panel) and C$^{18}$O (blue squares), obtained through Gaussian fitting. In the figure, a large-scale, sine-shaped velocity oscillation can be found in the velocity distribution of both $^{13}$CO and C$^{18}$O toward the central part of the major branch of filaments where three massive clumps are located, which has a good correspondence in the $N_{\rm H_2}$ distribution. This periodic velocity oscillation can be fitted as the sine function defined by an amplitude and a wavelength. The two parameters in sequence are $\sim$0.38 km s$^{-1}$ and $\sim$3.6\,pc for
$^{13}$CO, and $\sim$0.31 km s$^{-1}$ and $\sim$3.4\,pc for C$^{18}$O. The similar results derived from both species indicate that the velocity distributions along the center part of the major branch of the filaments are similar in both  $^{13}$CO and C$^{18}$O emission.

The similar pattern of density and velocity oscillations has been reported in Taurus/L1517 by \citet{2011A&A...533A..34H} and in the G350.54+0.69 filamentary cloud by \citet{2019MNRAS.487.1259L}.
Accordingly, two models of the kinematics of clump/core formation \citep{2011A&A...533A..34H} and the large-scale external physical perturbations (e.g.,  magnetohydrodynamic (MHD) transverse wave \citealt{2019MNRAS.487.1259L}) were proposed. In the core formation model, the motions of clump/core-forming gas converge toward the clump/core centre along the filament, leading to a $\lambda$/4 shift between the sinusoidal perturbations of density and velocity as characterised by the enhanced density peaking at the position of vanishing velocity (see \citealt{2011A&A...533A..34H} for illustration). For the other model, the large-scale external perturbation oscillates toward and away from us while propagating along the filament, leading to the observed velocity oscillation being blue- and red- shifted with respect to the systemic velocity, respectively. Ideally, the peaks of density enhancements and the velocity extremes (i.e. red/blue-shifted velocities) are aligned with one another if the filament (density) oscillates sinusoidally (see \citealt{2019MNRAS.487.1259L} for more details).

However, the second model is applicable to straight filamentary structures. The filamentary structure in G326 is curved, with the filaments on both sides of the C2 clump position almost at a 90-degree angle. Therefore, the large-scale external physical perturbations model is not suitable for explaining the oscillations observed in the filamentary structure in the G326 region. As shown in Fig.~\ref{fig:velocity_oscillations}, we do not observe the $\lambda$/4 shift between the velocity and the density patterns, nor the correspondence between the velocity extremes and density enhancements for all three clumps. Instead, the density peaks of C1 and C3 are found close to the position of vanishing velocity shown by a horizontal dashed line, while the density peak of C2 is near the velocity extreme. The results are close to the clump/core formation model, and the absence of an observed $\lambda$/4 shift between the velocity and density patterns near the C2 clump might be attributed to two factors: (1) C2 is situated at the central hub of the filamentary structure, where multiple filamentary structures from different directions transport material, leading to complex velocities at the C2 position. (2) We can observe that the two main filamentary structures on either side of the C2 clump form an almost 90-degree angle. We cannot precisely determine their true spatial positions, and the orientation of their spatial locations could also impact the velocity information near C2. If the orientations of the structures on both sides are consistent, it would result in the velocities on both sides of the C2 clump deviating from the position of vanishing velocity shown by a horizontal dashed line represented in Fig.~\ref{fig:velocity_oscillations}. We therefore suggest that the effect of clump formation would be responsible for the periodic oscillation in both velocity and density distributions observed here.

\subsection{High-mass star formation in the G326 cloud} \label{subsec:global_collapse_of_the_cloud}
\begin{figure}[ht!]
\begin{center}
\includegraphics[scale=1.7,angle=0]{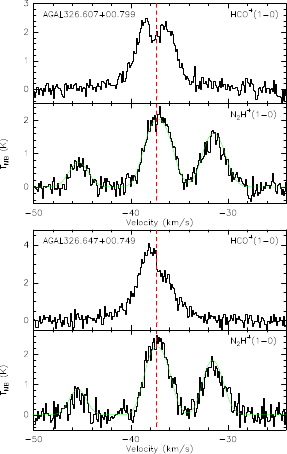}
\caption{Mean spectra of HCO$^{+}$(J = 1--0) and N$_{2}$H$^{+}$(J = 1--0) across two dense clumps of C2 (top panel) and C3 (bottom panel). The N$_{2}$H$^{+}$(J = 1--0) hfs fit is shown in green. The dashed red lines on the profiles indicate the peak velocities of N$_{2}$H$^{+}$(J = 1--0).}
\label{fig:clump_infall}
\end{center}
\end{figure}

The G326 cloud is undergoing star formation as inferred from the presence of several mid-infrared (i.e., 8 $\mu$m, see Fig.~\ref{fig:morphology_of_the_source}\,a) and far-infrared (i.e., 70 $\mu$m, see Fig.~\ref{fig:morphology_of_the_source}\,b) bright point sources embedded in the major branch of the filaments. Particularly, the global filament collapse (see above) toward the massive C2 and C3 clumps offers good conditions for higher mass star formation therein by transporting sufficient mass supply from ambient environment along the filaments to the clumps. The good association of the brightest 70\,$\mu$m point sources with the C2 and C3 clumps do favor higher-mass star formation taking place therein. Ongoing star formation could be signaled by the related infall motions \citep{2015MNRAS.450.1926H}. We therefore analyze the infall signatures associated with them to evaluate the star formation nature of the massive clumps of the cloud. Actually, the C1 clump has been identified as an infall candidate with a mass accretion rate of 2.44$\times$10$^{-3}$\,$M_{\odot}$ yr$^{-1}$ by \citet{2016MNRAS.461.2288H}. Following the same approach as in \citet{2016MNRAS.461.2288H}, we attempted to identify the infall signatures associated with the C2 and C3 clumps. The mapping data sets of both optically thin N$_{2}$H$^{+}$~(J = 1--0) and optically thick HCO$^{+}$~(J = 1--0) lines were obtained from the Millimetre Astronomy Legacy Team 90 GHz (MALT90) Survey \citep{2013PASA...30...57J} that is public for use. The clump-average spectra of both HCO$^{+}$ and N$_{2}$H$^{+}$ for the C2 and C3 clumps are shown in Fig.~\ref{fig:clump_infall}. Here, infall motions can be derived from the optically thick HCO$^{+}$~(J = 1--0) line, characterised either by double peaks with a brighter blue one, or by a skewed blue peak, in both of which the optically-thin N$_{2}$H$^{+}$~(J = 1--0) line should peak at the dip of the optically-thick line \citep{2015MNRAS.450.1926H,2016MNRAS.461.2288H,2021ApJS..253....2H}. As a result, the average spectrum of HCO$^{+}$ satisfies the characteristics above mentioned for both C2 and C3, lending a support for the infall motions ongoing therein.

Following \citet{2016MNRAS.461.2288H}, the clump-scale mass accretion rate can be roughly estimated from the formulae: $\dot{M}$$_{inf}$ =4$\pi$$R^{2}$$V_{\rm inf}$$n$ \citep[eq.\,5 of][]{2010A&A...517A..66L}, where $V_{\rm inf}$ = $V_{\rm N_{2}H^{+}}$ $-$ $V_{\rm HCO^{+}}$ is an estimate of the infall velocity, $n$=$M_{\rm clump}$/(4/3$\pi$$R^{3}$) is the average clump volume density, and $R$ is the effective radius of the clump extracted from \citet{2014A&A...568A..41U}. The spectrum of N$_{2}$H$^{+}$(J = 1--0) with a hfs was fitted using the hfs method in \emph{CLASS} to obtain the peak velocity ($V_{\rm N_{2}H^{+}}$ = $-$37.4 km s$^{-1}$ for C2 and C3). In Fig.~\ref{fig:clump_infall}, the hfs fitting curves are shown in green solid lines and the peak velocities of N$_{2}$H$^{+}$(J = 1--0) are indicated by red dashed vertical lines. $V_{\rm HCO^{+}}$ is the velocity of the blue peak (i.e., the left peak, $V_{\rm HCO^{+}}$ = $-$38.6 km s$^{-1}$, $V_{\rm inf}$ = 1.2 km s$^{-1}$), and the skewed peak ($V_{\rm HCO^{+}}$ = $-$38.1 km s$^{-1}$, $V_{\rm inf}$ = 0.7 km s$^{-1}$) for C2 and C3, respectively. Accordingly, the mass accretion rates are derived to be $\sim 5.8\times10^3$, and $\sim 3.1\times10^{-3}$\,$M_{\odot}$ yr$^{-1}$ typical of massive clumps where high-stars form \egcite{2015MNRAS.450.1926H,2016MNRAS.461.2288H}.

Furthermore, we have conducted estimations for the timescale of global longitudinal collapse. To achieve this, we employed the improved relation proposed by \citet{2015MNRAS.449.1819C} for estimating the collapse timescale based on the aspect ratio of the filamentary structure (A$_{0}$$>$2): $t_{\rm COL} \sim (0.49 + 0.26A_{0}) (G\rho_{0})^{-1/2}$. Within this equation, the density $\rho_{0}$ (defined as $n_{0}$ $\times$ m$_{\rm H_2}$, where m$_{\rm H_2}$ represents the mass of a hydrogen molecule) is selected using the same initial hydrogen number density $n_{0}$ = 4$\times$10$^4$ cm$^{-3}$, as utilized in the study by \citet{2014A&A...561A..83P}. The resulting collapse timescale is approximately 1.6 Myr$^{-1}$, consistent with the estimated collapse timescale for SDC13 as determined by \citet{2014A&A...561A..83P}. This suggests that within the collapse timescale in the G326 region, the three clumps, C1, C2, and C3, could accumulate sufficient mass to eventually form massive stars or star clusters.

To further study the relationship between the material transport rate along filamentary structures and the material infall rate of clumps on these structures, we adopt the mass transport rate relation along filaments proposed by \citet{2014A&A...561A..83P}: $\dot{M}_{\rm inf}=\pi (W/2)^2\rho_{\rm fil}V_{\rm inf}$, where W is the width of the filamentary structure, $\rho_{\rm fil}$ is the density of the filamentary structure, and $V_{\rm inf}$ is the material transport velocity. Here, our analysis is focused on the mass transport rate of the major branch of filaments (F1+F2). The average width of F1+F2 is taken to be approximately $W$ $\simeq$ 0.7 pc. The average material transport velocity at the positions of the clumps, C2 and C3, is estimated to be $V_{\rm inf}$ $\simeq$ 0.39 km s$^{-1}$. This leads to a calculated $\dot{M}_{\rm inf}$ $\simeq$ 1.73$\times$10$^{-4}$\,$M_{\odot}$ yr$^{-1}$. We find that the mass transport rate along the filamentary structure is an order of magnitude lower than the mass infall rate of the clumps (C1--C3). Even considering that clumps located at the junction of multiple filamentary structures could experience material transport from several filaments, the mass transport rate along the filamentary structure remains lower than the clump's mass infall rate. This discrepancy arises due to the fact that the density $\rho_{\rm fil}$ employed in calculating the mass transport rate represents an average density spanning the entire filamentary structure. The resulting $\dot{M}_{\rm inf}$ represents an average mass transport rate over the entire filamentary structure, rather than the true mass transport rate near the clumps. When taking into account the material transport rate near the clumps, with $V_{\rm inf}$ $\simeq$ 0.39 km s$^{-1}$, this velocity, although lower than the clump's infall velocity, is of a comparable magnitude. Taking into account that the density of the filamentary structure near the clumps is close to that of the clumps, and each clump is supplied material by at least two filamentary structures, we can further infer that the material transport along the filamentary structure is sufficient to sustain the continuous material supply for clump collapse.

\section{Summary} \label{sec:sum}
We have presented the new APEX observations of  both $^{13}$CO  and C$^{18}$O (J = 2--1) toward
the G326 HFS cloud. With the ancillary data from the MALT90 (i.e., HCO$^{+}$(1--0) and N$_{2}$H$^{+}$(1--0) lines) and the Hi-GAL (70--500\,$\mu$m continuum) surveys, the new observations provide insights onto the dynamics related to star formation from cloud to clump scales. The major results are summarized as follows:

(i) The G326 HFS cloud constitutes a central hub and at least four hub-composing filaments. They can be grouped into a major branch of filaments (F1, and F2) and a side branch (F3--F5).
 The $\sim$111.3\,pc-long major branch has a total mass of $\sim5\times10^3$ $M_{\odot}$, and a mean temperature of $\sim$16.9 K, and contains three massive dense clumps (i.e., 370--1100 $M_{\odot}$ and 0.14--0.16 g~cm$^{-2}$ for C1--C3), one of them (i.e., C2) also coincided with the central hub.

 (ii) Filament-aligned velocity gradients are found along the five filaments. The ones along F1 and F4--F5 appear toward the C2 clump, while the one along F2 points toward C3. F3, on the other hand, does not directly feed the C2 or C3 clumps but rather distributes material to both the C2 and C3 clumps simultaneously. Additionally, we have identified a new filamentary molecular emission branch situated north of the C3 clump, likely directly supplying material to the C3 clump. These velocity gradients could be attributed to the gravitational collapse along the filaments toward the central hub (i.e., C2).

 (iii) The spacing between the compact cores within the major branch of filaments matches well with the observed core separation of 0.6 pc, further supporting the model of self-gravitating fluid cylinders fragmenting due to the "sausage" instability.

(iv) A periodic velocity oscillation along the major branch of filaments is revealed in both $^{13}$CO and  C$^{18}$O (2--1) emission with a characteristic wavelength of $\sim$3.5\,pc and an amplitude of $\sim$0.31--0.38\,km s$^{-1}$. In G326, this pattern of the velocity oscillation could  result from the clump-forming gas motions induced by gravitational instability.

(v) The clump-scale dynamics is represented by the infall signatures of the three massive clumps. Along with satisfying the mass-size relationship for high-mass star formation, the associated high mass infalling rates ($>10^3$\,$M_{\odot}$ Myr$^{-1}$) support the occurrence of high-mass star formation therein. Accordingly, the presence of the associated YSOs reflected from 70\,$\mu$m emission suggests that G326 is an HFS cloud of ongoing high-mass star formation. Furthermore, we infer that the transport of material along the filamentary structures is sufficient to sustain a continuous supply of material during the clump collapse process.

\acknowledgments
We would like to thank the referee for the report that contributed to improve the quality of this paper. This work was mainly funded by the Chinese Academy of Sciences (CAS) "Light of West China" Program under grant No. 2020-XBQNXZ-017. It was also partially funded by the Xinjiang Key Laboratory of Radio Astrophysics under Grant No. 2023D04033, the National Natural Science Foundation of China (NSFC) under grant 12103045, the National Key R\&D Program of China under grant 2022YFA1603103, the NSFC under grants 11973076, 12173075, the Regional Collaborative Innovation Project of Xinjiang Uyghur Autonomous Region under grant 2022E01050, and the Science Committee of the Ministry of Science and Higher Education of the Republic of Kazakhstan under grant AP13067768.

This research has made use of the data products from the MALT90 survey, the SIMBAD data base, operated at CDS, Strasbourg, France, and the data from \emph{Herschel}, a European Space Agency space observatory with science instruments provided by European led consortia.

%

\vspace{5mm}
\facilities{APEX, \emph{Herschel}, Mopra}


\software{GILDAS/CLASS \citep{2005sf2a.conf..721P,2013ascl.soft05010G}, Matplotlib \citep{2007CSE.....9...90H}, astropy \citep{2013A&A...558A..33A}}


\clearpage
\appendix{}
\restartappendixnumbering

\section{Estimating $^{13}$CO and C$^{18}$O column densities} \label{sec:CO_column_densities}
Assuming local thermodynamic equilibrium (LTE) and a beam filling factor of 1, we estimated the column densities of $^{13}$CO and C$^{18}$O pixel-by-pixel from the J = 2--1 emission. The optical depths
($\tau_{\rm ^{13}CO}$ and $\tau_{\rm C^{18}O}$) and column densities ($N_{\rm ^{13}CO}$ and $N_{\rm C^{18}O}$) were derived using the following equations:
\begin{equation}
  \tau_{\rm ^{13}CO} = - ln\left(1 - \frac{T_{\rm mb}({\rm ^{13}CO})}{10.58\left[\frac{1}{e^{10.58/T_{\rm ex}}-1} - 0.021\right]}\right)
\label{tau13}
\end{equation}

\begin{equation}
{\rm N_{^{13}CO}} = 1.257 \times 10^{14}~e^{\frac{5.29}{T_{\rm ex}}}\frac{T_{\rm ex} + 0.88}{1 - e^{\frac{-10.58}{T_{\rm ex}}}}
\int{\tau_{\rm ^{13}CO}{\rm dv}}
\label{N13}
\end{equation}
with
\begin{equation}
\int{\tau_{\rm ^{13}CO}{\rm dv}} = \frac{1}{J(T_{\rm ex}) - 0.222} \frac{\tau_{\rm ^{13}CO}}{1-e^{-\tau_{\rm ^{13}CO}}}
\int{T_{\rm mb}({\rm ^{13}CO}){\rm dv}}
\label{integ13}
\end{equation}
\begin{equation}
\tau_{\rm C^{18}O} = - ln\left(1 - \frac{T_{\rm mb}({\rm C^{18}O})}{10.54\left[\frac{1}{e^{10.54/T_{\rm ex}}-1} - 0.021\right]}\right)
\label{tau18}
\end{equation}
\begin{equation}
{\rm N_{C^{18}O}} = 1.262 \times 10^{14}~e^{\frac{5.27}{T_{\rm ex}}}\frac{T_{\rm ex} + 0.88}{1 - e^{\frac{-10.54}{T_{\rm ex}}}}
\int{\tau_{\rm C^{18}O}{\rm dv}}
\label{N18}
\end{equation}
with
\begin{equation}
\int{\tau_{\rm C^{18}O}{\rm dv}} = \frac{1}{J(T_{\rm ex}) - 0.221} \frac{\tau_{\rm C^{18}O}}{1-e^{-\tau_{\rm C^{18}O}}}
\int{T_{\rm mb}({\rm C^{18}O}){\rm dv}}
\label{integ18}
\end{equation}
The $J(T_{ex})$ parameter is $\frac{10.58}{exp(\frac{10.58}{T_{\rm ex}}) - 1}$ in the case of Eq.\,(\ref{integ13}) and $\frac{10.54}{exp(\frac{10.54}{T_{\rm ex}}) - 1}$ in Eq.\,(\ref{integ18}).
In all equations $T_{\rm mb}$ is the peak main brightness temperature and $T_{\rm ex}$ the excitation temperature. Under the assumption of LTE, all rotational levels of $^{13}$CO and C$^{18}$O molecules are populated according to the same excitation temperature ($T_{\rm ex}$). Here, we assumed that $T_{\rm ex}$ is equal to the dust temperature ($T_{\rm d}$).

\section{FWHM velocity width distribution of $^{13}$CO and C$^{18}$O for G326} \label{sec:FWHM_velocity_width_distribution}

\begin{figure*}[ht!]
\begin{center}
\includegraphics[scale=1.2,angle=0]{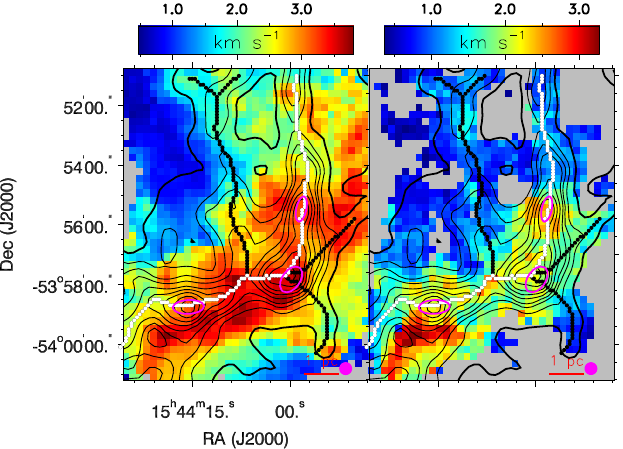}
\caption{FWHM velocity width distribution of $^{13}$CO (left panel) and C$^{18}$O (right panel) for G326, which derived from the moment method. Contours are the H$_{2}$ column density in levels of (0.9, 1.1, 1.3, 1.5, 1.9, 2.4, 3.0, 4.0, 5.0) $\times$ 10$^{22}$ cm$^{-2}$. Magenta ellipses mark the dust clumps. The main and branched skeletons are shown as white and black curves, respectively. The magenta filled circle in the lower right corner indicates the beam size of the CO observation by APEX.}
\label{fig:FWHM_velocity_width_distribution}
\end{center}
\end{figure*}

\bibliography{sample63}{}
\bibliographystyle{aasjournal}

\end{document}